%% file: main.tex
\documentclass[aps,prd,reprint,superscriptaddress,floatfix,altaffilletter,nofootinbib]{revtex4-2} 

\usepackage{graphicx}
\usepackage{dcolumn}
\usepackage{amsmath}
\usepackage{amssymb}
\usepackage{epstopdf}

\usepackage{color}
\usepackage{upgreek}
\usepackage{url}
\usepackage{here}

\usepackage{orcidlink}
\usepackage{siunitx}
\usepackage{xspace, tabularx}

\usepackage{multirow}
\usepackage{subcaption}

\makeatletter
\def\btt#1{\texttt{\@backslashchar#1}}%
\DeclareRobustCommand\bblash{\btt{\@backslashchar}}%
\makeatother

\newcommand{\todo}[1]{}
\renewcommand{\todo}[1]{{\color{red} TODO: {#1}}}

\begin{document}
\title{Measurement of the solar neutrino interaction rate below 3.49 MeV in Super-Kamiokande-IV}
\input{authors}
\date{June 1, 2026}

\input{commands}

\begin{abstract}
Super-Kamiokande (SK) has observed $^8$B solar neutrino elastic scattering at recoil electron kinetic energies ($E_{\text{kin}}$) as low as \SI{3.49}{MeV} to study neutrino flavor conversion within the Sun. At SK-observable energies, these conversions are dominated by the Mikheyev–Smirnov–Wolfenstein (MSW) effect. An upturn in the electron neutrino survival probability in which vacuum neutrino oscillations become dominant is predicted to occur at lower energies, but radioactive background increases exponentially with decreasing energy. New machine learning approaches provide substantial background reduction below \SI{3.49}{MeV} such that statistical extraction of solar neutrino interactions becomes feasible. This article presents an analysis of the solar neutrino interaction rate at $E_{\text{kin}} < \SI{3.49}{MeV}$ with the full SK-IV period, using data from a wideband intelligent trigger when available and with a boosted decision tree for event selection. A solar neutrino signal is observed between $\SI{2.99}{MeV} < E_{\text{kin}} < \SI{3.49}{MeV}$ with $2.76\sigma$ significance and a data to unoscillated Monte Carlo ratio of $0.307^{+0.112}_{-0.111}$. These additional low-energy data have a negligible effect on the $1\sigma$ intervals of the fits to the solar neutrino energy spectrum but has a noticeable effect on the best fit when using the exponential parametrization.
\end{abstract}

\maketitle


\section{\label{sec:intro}Introduction}
 Various experiments have observed solar neutrinos produced through the fusion chain in the core of the Sun and have confirmed neutrino flavor conversion through \nue disappearance \cite{solar_neutrinos,kii,ga71,gallex,gno,ski_2001,sno_2001,sno_nc}. Measurements from Super-Kamiokande (SK), Sudbury Neutrino Observatory (SNO), and KamLAND of the solar neutrino interaction rate and corresponding \nue survival probability ($P_{ee}$) are consistent with the predictions of the Mikheyev–Smirnov–Wolfenstein (MSW) effect in which \nue undergo an adiabatic conversion as they pass through a high-mass-density region and convert into \nutwo \cite{ms, w, skiv, sno, kamland_b8}, while measurements from Borexino and KamLAND of lower-energy \textit{pp}, $^7$Be, and \textit{pep} neutrinos are consistent with the usual averaged vacuum oscillations \cite{borexino, kamland_be7}. An ``upturn'' in $P_{ee}$ is therefore expected to occur between ${\approx}1$ and $\SI{\approx7}{MeV}$ in neutrino energy as flavor conversions transition from the MSW-dominant region with $P_{ee} \sim \cos^4{\theta_{13}}\sin^2{\theta_{12}}+\sin^4{\theta_{13}}$ at higher energies to the vacuum oscillation-dominant region with $P_{ee} \sim \cos^4{\theta_{13}}(1-\frac{1}{2}\sin^2{2\theta_{12}})+\sin^4{\theta_{13}}$ at lower energies \cite{quasivacuum}. A measurement of $P_{ee}$ in this region would be an important test of the MSW effect and electroweak physics in general. However, high rates of radioactive background radiation in this transition region have prevented observation of the solar upturn.

SK has published a solar analysis using the full fourth data period of SK running (SK-IV) with 2970 days of live time \cite{skiv} for a total of 5805 days across all phases. Due to the low background levels in SK-IV and the large dataset size, this phase provides the bulk of the sensitivity for the SK solar analysis and has the lowest energy threshold: \SI{4}{MeV} in recoil electron \textit{total} energy (\SI{3.49}{MeV} \textit{kinetic}). In this analysis, SK favors the existence of an upturn by $1.2\sigma$. Together with SNO, the significance increases to $2.1\sigma$ \cite{skiv}.

Toward the end of SK-IV, several efforts were made to increase the trigger efficiency at low energies. These include lowering the triggering threshold of the simple coincidence trigger in the data acquisition (DAQ) and introducing an entirely separate trigger system known as the ``wideband intelligent trigger'' (WIT) running in parallel. The purpose of the analysis presented in this article is to lower the energy threshold of the SK-IV solar analysis by using improved event selection procedures and by making use of WIT data when available.

\section{Super-Kamiokande-IV}\label{sec:sk}
Super-Kamiokande is a \SI{50}{kton} cylindrical water Cherenkov detector in Gifu, Japan with \SI{2700}{m} water-equivalent rock overburden \cite{sk}. In SK-IV, the \SI{32.5}{kton} inner detector is surrounded by 11,129 \SI{50}{cm} photomultiplier tubes (PMTs) that detect Cherenkov light emitted by particles traveling through the detector volume with energies above the Cherenkov threshold. 1885 additional \SI{20}{cm} PMTs are used for the separate outer detector, which is used to reject events originating outside of the inner detector volume. SK-IV ran from October 2008 to May 2018 with 2970 days of live time. Aside from being the longest of all SK phases, SK-IV is characterized by improved front-end electronics \cite{qbee} and low levels of radioactivity compared to previous phases.

SK observes solar neutrinos through elastic scattering on electrons in which the recoil electron produces a single \ang{42} Cherenkov cone. All event reconstruction for the solar analysis assumes only a single electron ring is present, and methods for evaluating the hit pattern's consistency with a single ring are important for background rejection. Elastic scattering preserves the directionality of the neutrino. Determining the number of solar neutrino events present in a selection is therefore done through fits to the solar angle distribution. The primary limit on electron direction resolution is multiple Coulomb scattering. This effect increases with decreasing energy and exceeds the impact of both the initial weak scattering angle and detector-related effects.

\begin{figure}[t]
\centering
\includegraphics[width=0.9\linewidth]{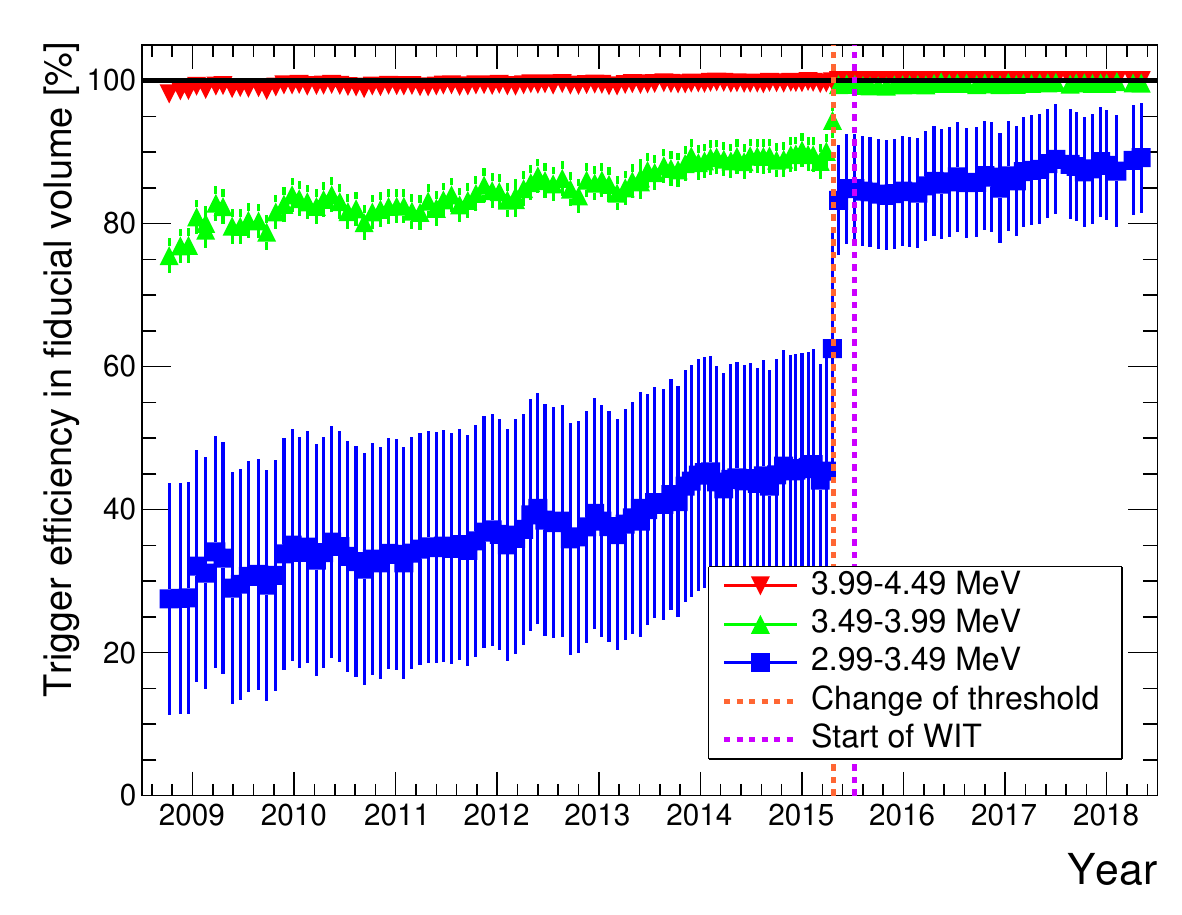}
\caption{\label{fig:trigger_vs_time} The time variation of the SLE trigger efficiency estimated by the MC simulation of $\mathrm{^{8}B}$ solar neutrinos shown as the average of 30-day intervals. The green and red plots represent the lowest two energy bins of \cite{skiv}, and the 2.99--\SI{3.49}{MeV} energy bin (blue) is discussed in this article. After changing the trigger threshold ~(orange dashed line) from 34 hits to 31 hits, trigger efficiency improved significantly. The start of data-taking with the separate WIT system ~(violet dashed line) is shown for reference.}
\end{figure}

Figure \ref{fig:trigger_vs_time} shows the trigger efficiency over time of the ``super low energy'' (SLE) triggers of the existing hit coincidence trigger system for the lowest energy bins. The gradual increase in trigger efficiency is driven by the improvement in PMT gain and water transparency throughout SK-IV \cite{skiv}. Following efforts to reduce the radon content in the purified water supply \cite{radon}, the SLE trigger threshold was reduced from 34 hits in \SI{200}{ns} to 31 in May 2015. This change in threshold brought the trigger efficiency at the \SI{3.49}{MeV} threshold to nearly $100\%$ within a \SI{22.5}{kton} fiducial volume defined as $\SI{>2}{m}$ from the inner detector walls.

\begin{figure}[t]
\centering
\includegraphics[width=0.9\linewidth]{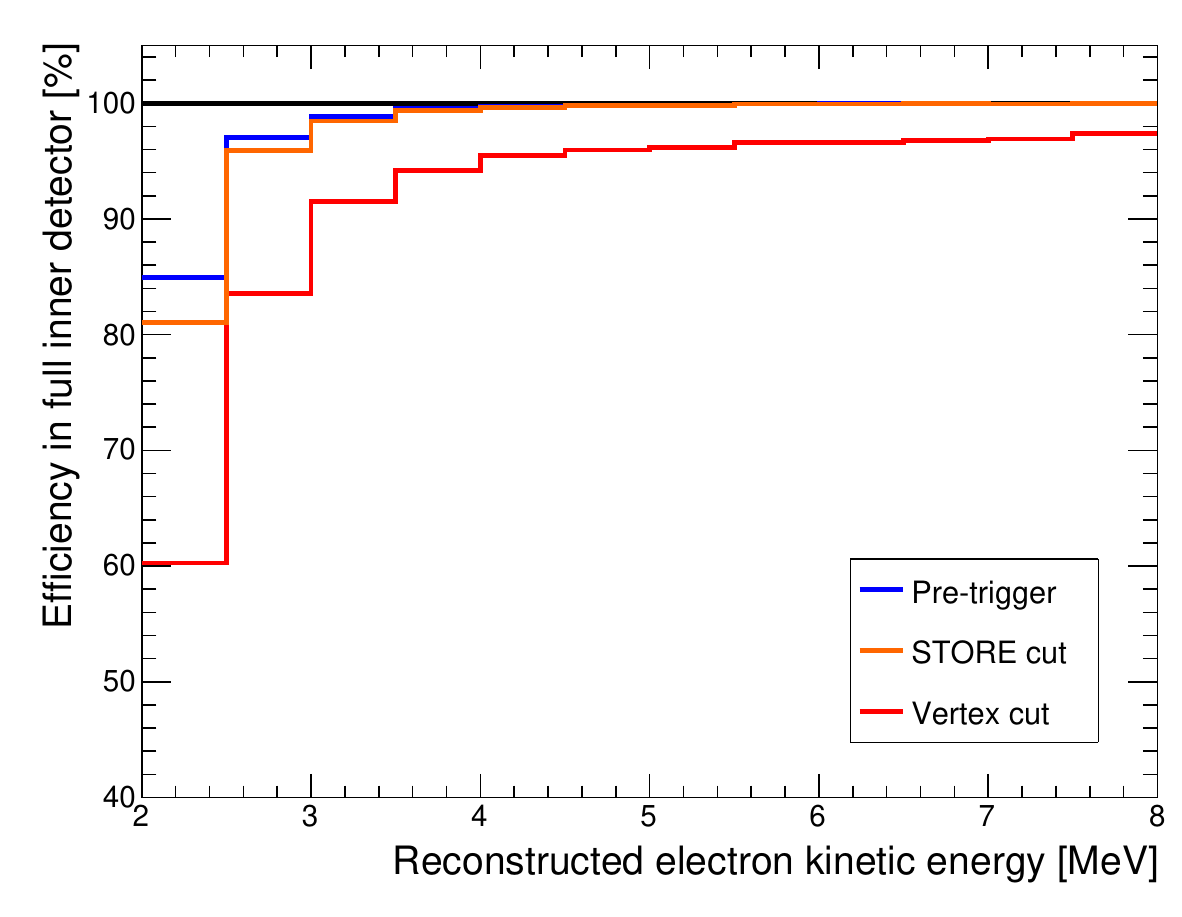}
\caption{\label{fig:wit_eff} The trigger efficiency of the WIT system as a function of reconstructed electron kinetic energy based on solar $\mathrm{^{8}B}$ MC. The blue, orange, and red plots show the cumulative efficiency following the pretrigger, the cut on the number of hits coincident with potential vertices, and the cut on the final vertex position, respectively.}
\end{figure}

The independent WIT system was installed soon after in July 2015 \cite{wit2015, wit2017}. In an alternative approach to improving trigger efficiency, WIT acts as a separate DAQ running in parallel to the hit coincidence trigger. The WIT system consists of a collection of high-performance computers (seven computers with 16 CPU cores each during SK-IV) and receives the same data as the hit coincidence trigger as unsorted \SI{23}{ms} blocks. With hyperthreading enabled, each physical CPU core can process two \SI{23}{ms} blocks in parallel with a trigger algorithm that applies a variable and much looser threshold of 11 hits above the current dark noise rate in \SI{230}{ns} (the ``pretrigger''). The efficiency of the pretrigger is near 100\% down to \SI{2.49}{MeV} kinetic energy (Fig. \ref{fig:wit_eff}) \cite{wit2017}. A second filter known as ``STORE'' then creates a list of potential vertex positions from the PMT four-hit combinations and requires that at least effectively 6.7 hits are coincident within \SI{5}{ns} after time-of-flight subtraction for at least one of the potential vertices \cite{spal_cut}. The process then conducts full vertex reconstruction with the ``BONSAI'' algorithm used in previous analyses \cite{skii, reco}. Events that have between 6.7 and 11 coincident hits from the STORE filter above are required to pass the \SI{2}{m} fiducial volume cut, and events with 11 or more coincident hits must simply be within the inner detector volume. Each block is then sent to a separate organizer machine that sorts the blocks in time. The WIT dataset had not previously been used as a source of signal data for the solar event selection, but it has been used in \cite{skiv} to reject low-energy neutron cloud events resulting from background cosmic ray muon spallation \cite{spal_cut}. In later phases, the WIT system has been used to serve as a trigger and raw data buffer for potential presupernova and supernova burst neutrinos \cite{preSN}.

\section{Sources of background}
By far the largest source of background below the analysis threshold in \SI{2.49}{MeV}--\SI{3.49}{MeV} is radioactive beta decay. In the WIT dataset after first reduction cuts, there are $O(10^7)$ radioactive background events per day compared to $O(10)$ solar neutrino interactions. The most dominant source of radioactive background is $^{214}$Bi beta decay with a Q-value of \SI{3.3}{MeV} resulting from the decay of radon gas present in the detector \cite{radon, buffer}. By controlling the supply water temperature, the water currents within the inner detector are designed in such a way as to collect as much of this radon as possible near the bottom of the detector \cite{radon, calib}. A $^{214}$Bi beta in the fiducial volume is indistinguishable from a solar neutrino recoil electron at the same energy. The next leading source of radioactive background is beta-gamma decay of $^{208}$Tl present in the PMT glass. This decay has a Q-value of \SI{5.0}{MeV} and most often includes a \SI{2.6}{MeV} gamma. These events therefore occur near the detector walls, but occasional coincidental triggers with dark noise hits can result in these events being reconstructed within the \SI{2}{m} fiducial volume. Similarly, although the energies of these events are below even those considered in this analysis, poor energy resolution and such coincidental triggers cause these decays to be the dominant background source as high as \SI{6}{MeV}.

The next leading source of background is again radioactive decay but caused by isotopes created during cosmic ray muon spallation \cite{spal}. While electromagnetic showers induced by cosmic ray muons are an important background at higher energies, the dominant spallation products at these low energies are radioactive isotopes created by the interaction of hadronic showers \cite{spal_cut} with $^{16}$O. The leading spallation-induced radioactive isotope is $^{16}$N, which is long-lived, lasting up to several seconds, and can be found meters away from the showering vertex before undergoing beta decay \cite{ski, banching_ratio}. There are $O(10^3)$ events per day due to the spallation isotope background.

Other sources of background are much more easily removed with existing cuts. These include events due to calibration sources and flashing PMTs.

\section{Event selection}

\begin{table*}[t]
\caption{\label{tab:vars}Variables input to the BDT.}
\begin{ruledtabular}
\begin{tabular}{lp{6.5in}}
Index & Description\\
\colrule
1 & Number of PMT hits in the event \\
2--3 & Reconstructed BONSAI vertex position in the detector as $r$ and $z$ \\
4--6 & Reconstructed direction as a unit vector $[dx, dy, dz]$\\
7 & Closest distance from the vertex to the inner detector wall\\
8--10 & Point of intersection with the wall along a line starting from the vertex aligned with the reverse direction vector as $[x, y, z]$\\
11 & Distance from the vertex to the wall along this line (effective wall distance)\\
12 & Distance from the vertex to the \SI{8.5}{kton} ``tight'' fiducial volume boundary (see \cite{skiv_2016}, Sec. III.A.4)\\
13 & Value of a new 2D polynomial fit to the background vertex positions in $r^2$ and $z$\\
14 & Reconstructed total event energy\\
15 & Vertex fit goodness, $g_t$, higher for smaller PMT hit timing residuals (see \cite{skii}, Sec. III.B)\\
16 & Hit pattern goodness, $g_p$, lower for better azimuthal symmetry of the Cherenkov cone (see \cite{skii}, Sec. III.B)\\
17 & $g_t^2 - g_p^2$\\
18 & ``Multiple scattering goodness'' (MSG), higher for better quality direction reconstruction associated with higher energy and less electron scattering (see \cite{skiv_2016}, Sec. II.C.4)\\
19 & Smallest radius containing one-fifth of the hit PMTs, smaller for events originating at or near PMT glass (see \cite{skiv_2016}, Sec. III.A.1)\\
\end{tabular}
\end{ruledtabular}
\end{table*}

The event rate of radioactive background increases exponentially with decreasing energy approaching the true energy of these betas and gammas. The background rejection of the solar $\mathrm{^{8}B}$ neutrino selection cuts used in previous SK solar analyses \cite{skiv, skiv_2016, skii} (the ``existing cuts'') is not sufficient to observe a signal below the previous \SI{3.49}{MeV} threshold. Several different event selection approaches were studied to improve background rejection in this energy region including several machine learning-based methods. These methods include convolutional neural networks trained on 2D event display images \cite{watchmal}, graph neural networks trained on PMT hit information in 3D, boosted decision trees (BDTs) trained on reconstructed variables, and hybrid approaches that make use of both PMT hit information and high-level variables. This analysis uses a BDT since its performance was found to match or exceed other methods and is less computationally intensive than other methods.

\begin{figure*}[t]
\centering
\begin{subfigure}{0.24\linewidth}
\centering
\includegraphics[width=1.0\linewidth]{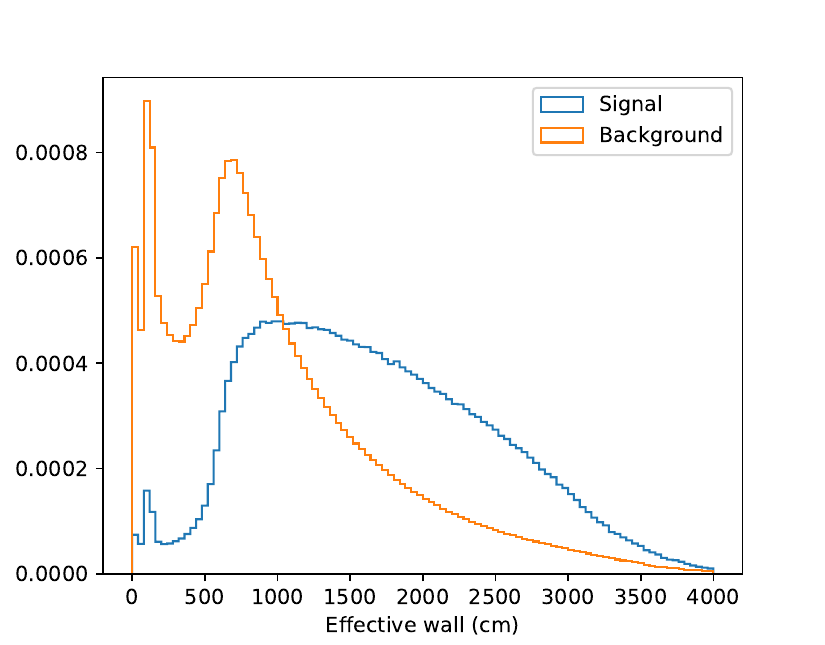}
\caption{11: Effective wall distance}
\end{subfigure}
\begin{subfigure}{0.24\linewidth}
\centering
\includegraphics[width=1.0\linewidth]{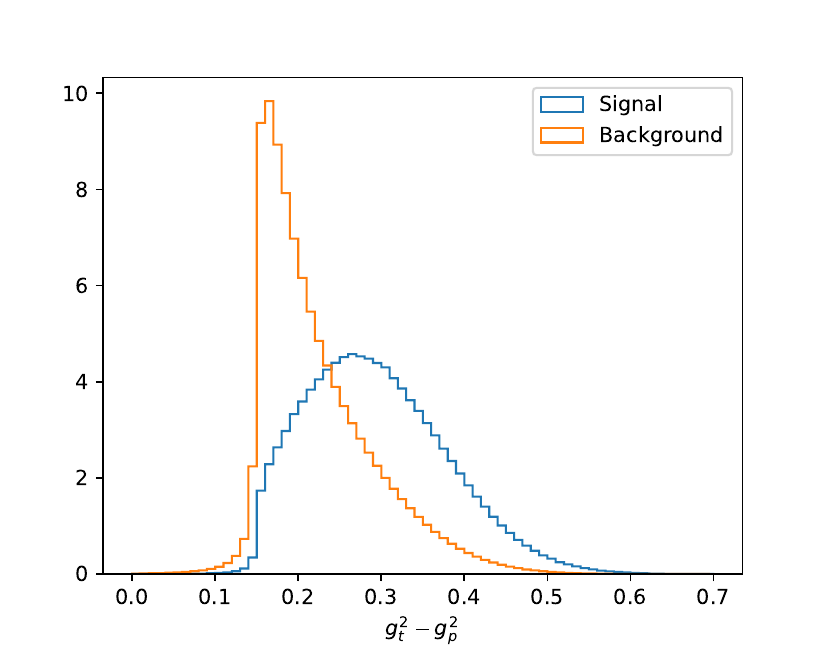}
\caption{17: $g_t^2 - g_p^2$ fit goodness}
\end{subfigure}
\begin{subfigure}{0.24\linewidth}
\centering
\includegraphics[width=1.0\linewidth]{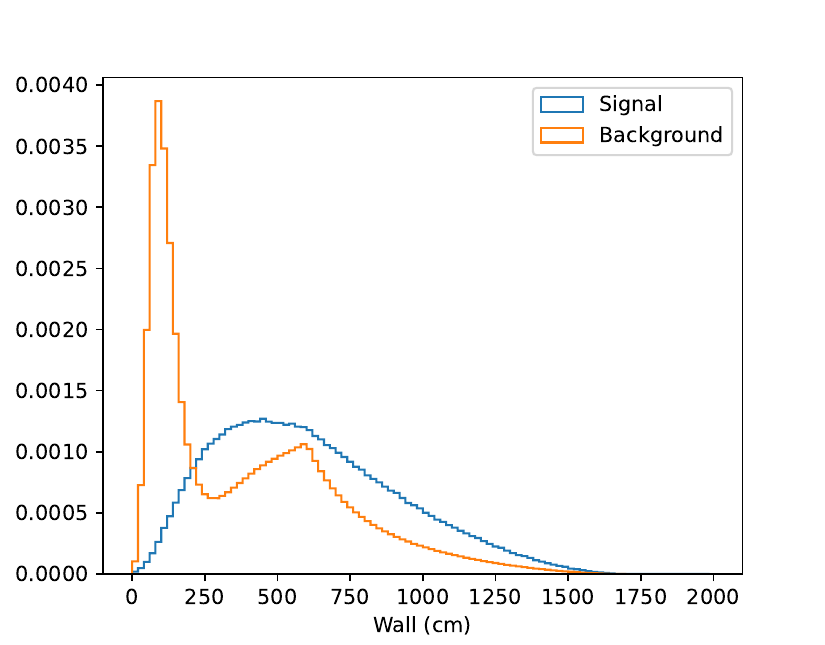}
\caption{7: Distance to wall}
\end{subfigure}
\begin{subfigure}{0.24\linewidth}
\centering
\includegraphics[width=1.0\linewidth]{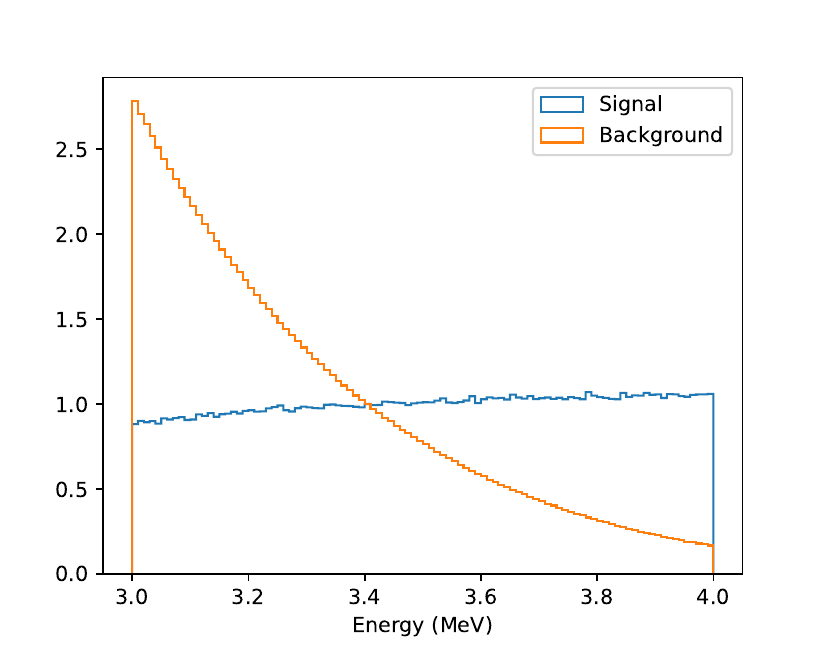}
\caption{14: Total energy}
\end{subfigure}
\caption{\label{fig:vars} Distributions of the most important reconstructed variable inputs to the BDT in the WIT dataset for signal (blue) and background (orange) following preselection cuts. The numbers refer to the indices in Table \ref{tab:vars}. Histograms are normalized to an area of 1. Variable importance is determined by the XGBoost ``Total gain'' metric \cite{xgboost}.}
\end{figure*}

\subsection{Boosted decision tree}

A BDT was trained to differentiate between solar neutrino signal events and radioactive background. For the signal Monte Carlo (MC) simulation, we follow the same procedure as in \cite{skiv}. Solar neutrino recoil electrons events are generated within the inner detector volume. For the purpose of generating a training dataset, however, the direction of the recoil electron was randomized drawing from a spherically uniform distribution. This approach prevents the BDT from potentially learning about the solar direction and causing biases in the solar angle distribution so that any intentional use of the solar direction can be reserved for the postselection solar angle fitting in Sec. \ref{sec:solfit}. Detector simulation is then performed with \textsc{Geant3} \cite{geant3} modified for SK-IV.

We generate two separate datasets corresponding to each of SLE and WIT data and train separate BDTs for each of these datasets. For the SLE dataset, the detector simulation is run with simulated dark noise. For the WIT dataset, dark noise simulation is turned off and events are instead overlaid onto ``dummy triggers'' containing only real dark noise hits. The respective trigger algorithm is then run on each of the datasets. Separate testing datasets are also generated in the same manner but without the recoil electron direction randomization so that these datasets respect the actual solar direction. These signal testing datasets are also used for the signal extraction step in Sec. \ref{sec:solfit}.

Following signal dataset generation but before training, a series of loose preselection cuts are applied. These cuts are chosen to exactly match those used on data as events failing these cuts were deleted. Table \ref{tab:vars} provides a list of reconstructed variables, the majority of which have been used for event selection in previous solar analyses. The preselection cuts are based on the four most significant variables for background rejection: the distance to wall, effective wall distance, $g_t^2 - g_p^2$, and, notably, the energy. This cut in kinetic energy is \SI{2.99}{MeV}, \SI{2.49}{MeV}, and \SI{2.24}{MeV} for the SLE 34-hit, SLE 31-hit, and WIT datasets, respectively. With the motivation of having the BDT perform the bulk of the event selection, these cuts are not tightened any further in order to include as much data as possible. Figure \ref{fig:vars} shows distributions of these variables for the WIT dataset following these preselection cuts. 

For the background sample, due to the abundance of radioactive background data and the extremely low signal-to-background ratio ($O(10^{-6})$) following only the loose preselection cuts, real data were used as a stand-in for proper background simulation. We assume that the BDT will not be largely influenced by the small fraction of real signal events that are labeled as background for training. This approach eliminates the need to consider systematic errors in background simulation, limiting such errors to signal simulation only. Here, 1\% of the available data was randomly selected to use as the BDT training dataset and another 1\% for the testing dataset.

\begin{figure}[t]
\centering
\includegraphics[width=0.9\linewidth]{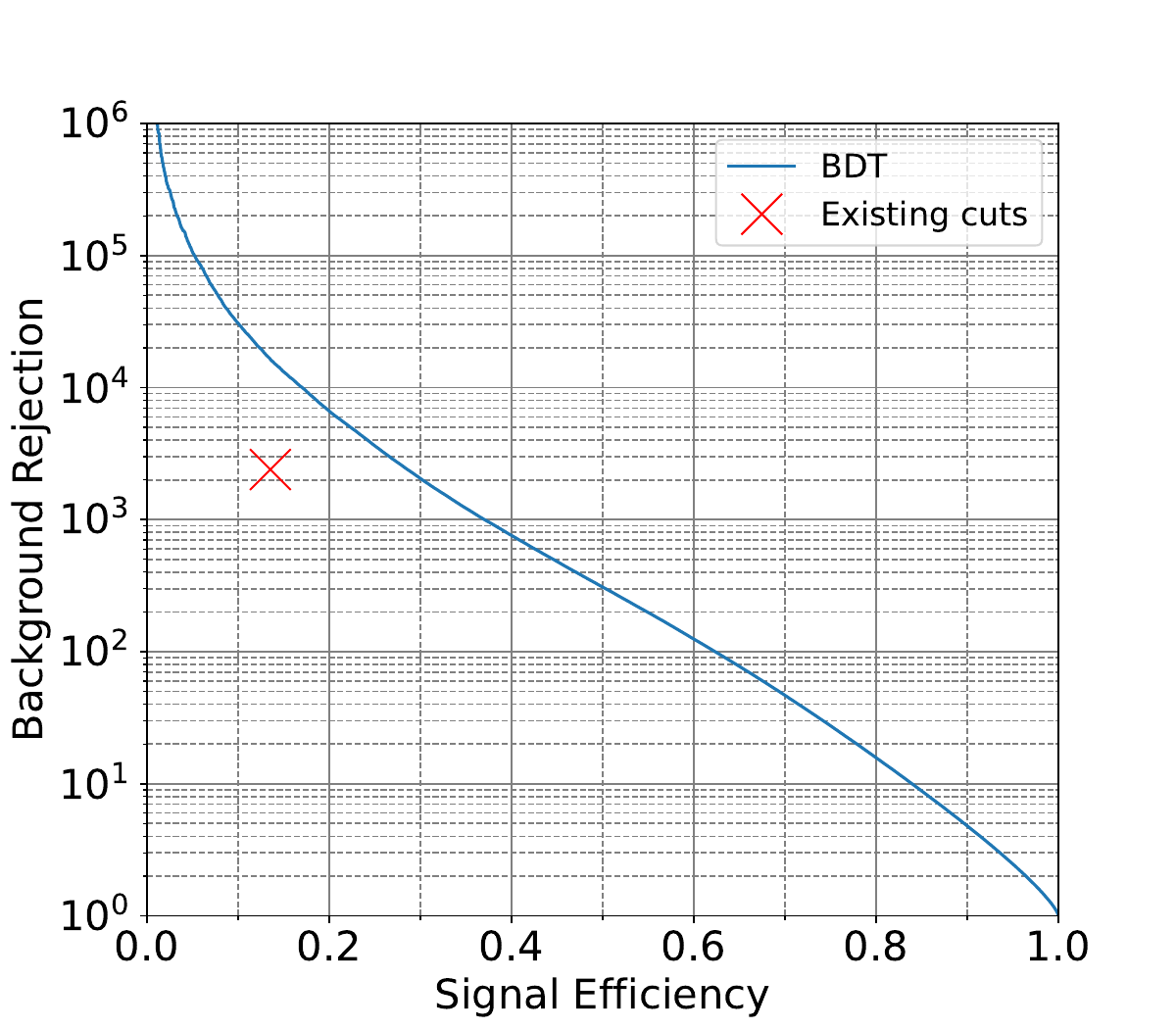}
\caption{\label{fig:roc} BDT background rejection calculated as the inverse of the background acceptance rate as a function of signal efficiency for the WIT dataset in \SI{2.49}{MeV}--\SI{3.49}{MeV}. The performance of the existing cuts used in \cite{skiv} except those described in Sec. \ref{sec:final_sel} is shown as a red $\times$.}
\end{figure}

The BDTs are trained on the 19 variables in Table \ref{tab:vars}. XGBoost v1.5.0 was used to train, test, and implement the BDTs \cite{xgboost}. Since the signal-to-background ratio is especially poor in this low-energy region, the training procedure was optimized for this high-rejection context. However, typical methods used to address ``class-imbalanced'' datasets including class reweighting and using alternative loss functions, such as the ``recall loss'' and ``focal loss'' \cite{imbalance, recall, focal}, were found to be less effective than the following method: three separate BDTs are trained in succession. The first is trained on a subset of the overall training sample with equal parts signal and background. The first BDT is then used to perform inference on the rest of the training dataset, and a loose cut on the BDT output score is applied, chosen to reduce background by a factor of 10. A second BDT is then trained on another balanced subset of surviving signal and background events using the same variables as in Table \ref{tab:vars}. This process is repeated to train a third model with only events that pass these cuts on the scores of the first two models. A completely separate hold-out test set is used to evaluate the model. Although each individual BDT is trained on a balanced dataset, this approach requires much more background than signal for training overall. Figure \ref{fig:roc} shows the selection performance of this BDT for the WIT dataset, which offers increased background rejection in \SI{2.49}{MeV}--\SI{3.49}{MeV} for the same signal efficiency as the existing cuts by a factor of 6.8.

\subsection{Final selection}\label{sec:final_sel}
Following the implementation of the respective BDTs on the SLE and WIT datasets, other selection cuts are applied that are intended to remove sources of background other than the dominating $^{214}$Bi and $^{208}$Tl decay background events. These include cuts to remove flashing PMTs, events near permanently installed calibration sources, bad runs, including a high background rate period near the end of SK-IV, and cosmic ray muon spallation, including a reoptimized spallation log-likelihood cut, multiple spallation cut, and neutron cloud cut \cite{spal_cut}.

The cut on the BDT output score is applied last, and the value of this cut is tuned to optimize the $S/\sqrt{B}$ figure of merit for signal MC $S$ and data $B$.

As WIT was implemented after the switch to the 31-hit trigger threshold, the WIT dataset does not overlap with the 34-hit SLE dataset. For the 31-hit period, only runs in which WIT data were not available were included. The live times of these three SK-IV datasets are summarized in Table \ref{tab:livetimes} and total 2970 days.

\begin{table}[b]
\caption{\label{tab:livetimes}Live times of the datasets used in this selection.}
\begin{ruledtabular}
\begin{tabular}{lccc}
Dataset & Start & End & Live time (days)\\
\colrule
34-hit SLE & 6 October 2008 & 1 May 2015 & 2047\\
31-hit SLE & 1 May 2015 & 30 May 2018 & 305\footnote{Only includes live time when WIT was not available.}\\
WIT & 10 July 2015 & 30 May 2018 & 618\\
\end{tabular}
\end{ruledtabular}
\end{table}

\begin{figure*}[t]
\centering
\begin{subfigure}{0.23\linewidth}
\centering
\includegraphics[width=1.0\linewidth]{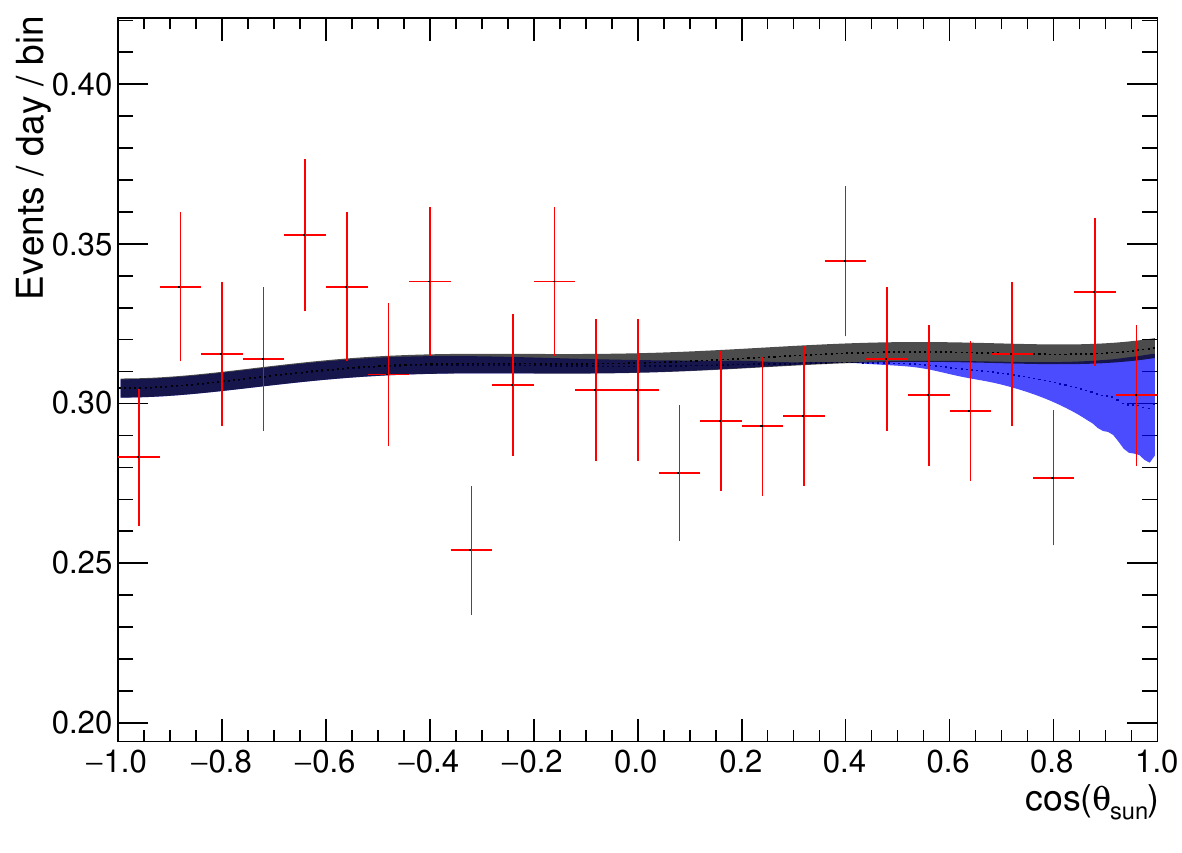}
\caption{\SI{2.49}{MeV}--\SI{2.99}{MeV}, MSG bin 1}
\end{subfigure}
\begin{subfigure}{0.23\linewidth}
\centering
\includegraphics[width=1.0\linewidth]{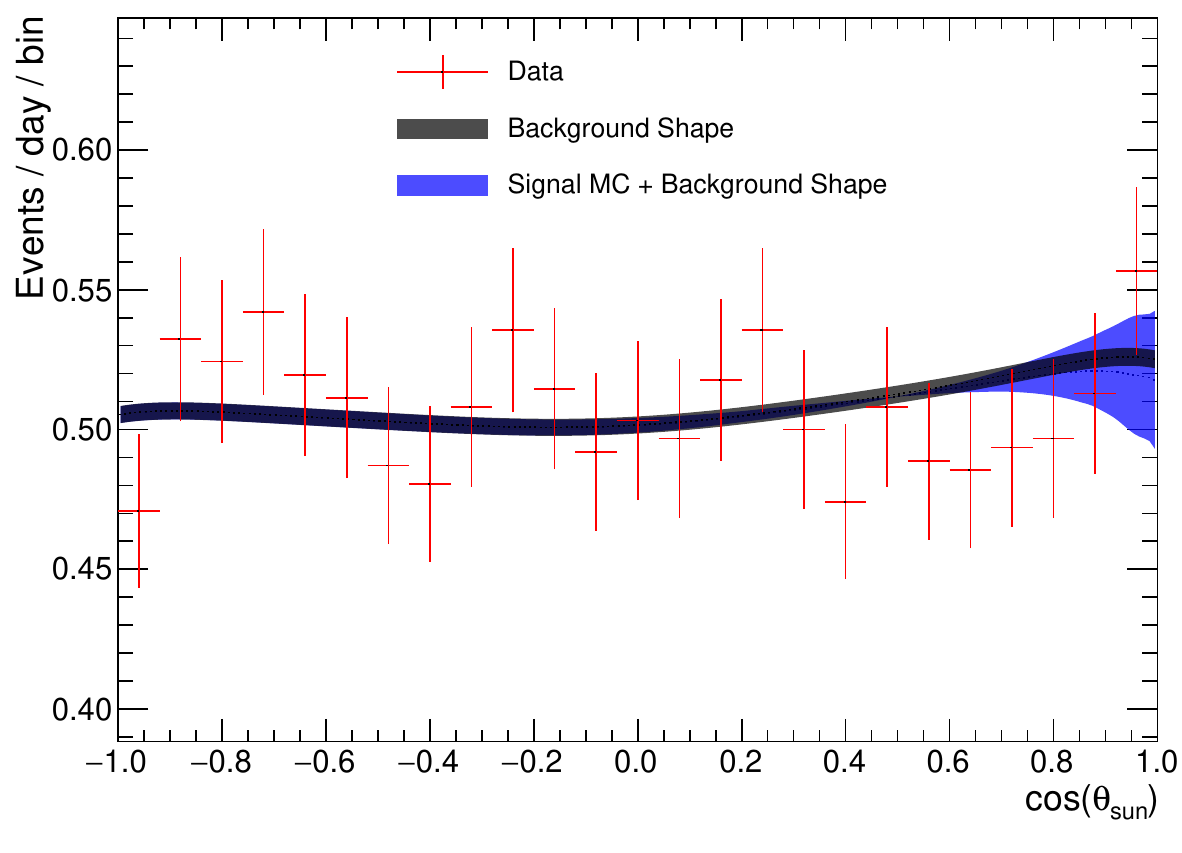}
\caption{\SI{2.49}{MeV}--\SI{2.99}{MeV}, MSG bin 2}
\end{subfigure}
\begin{subfigure}{0.23\linewidth}
\centering
\includegraphics[width=1.0\linewidth]{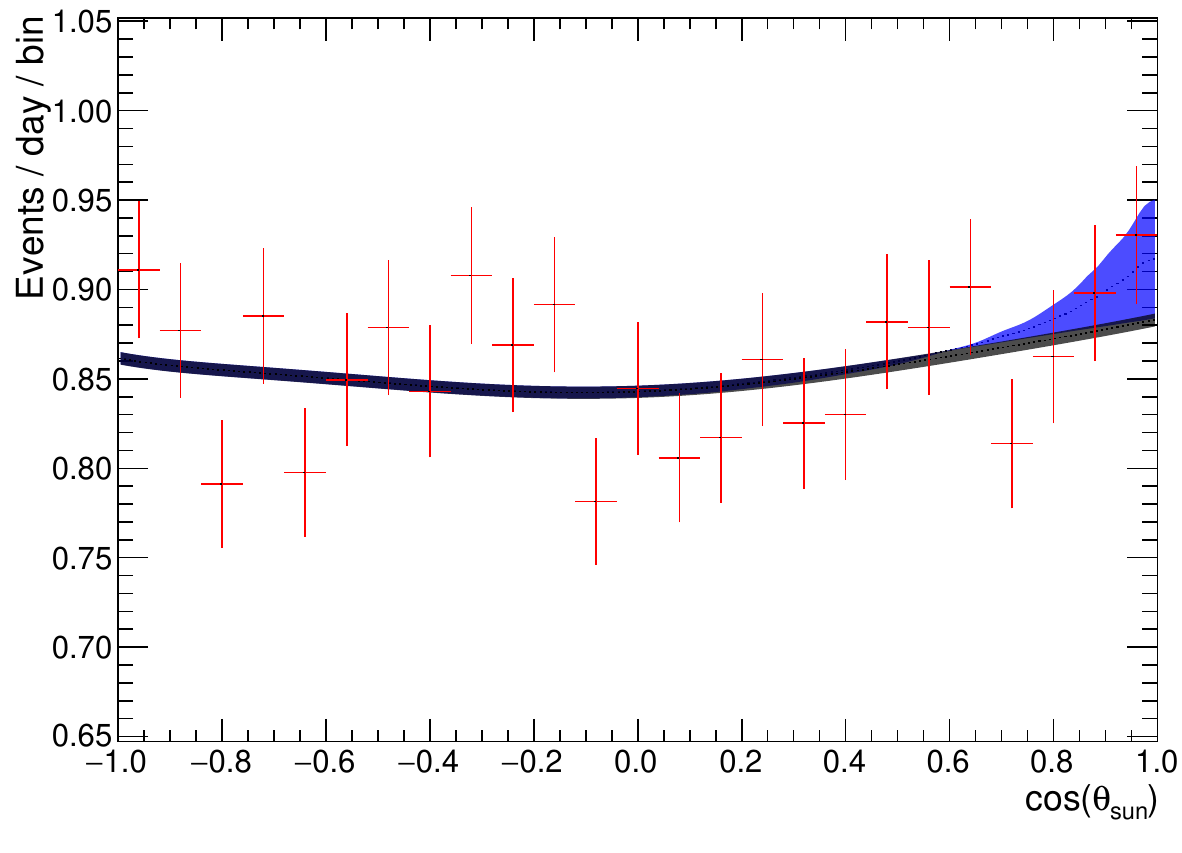}
\caption{\SI{2.49}{MeV}--\SI{2.99}{MeV}, MSG bin 3}
\end{subfigure}\\
\begin{subfigure}{0.23\linewidth}
\centering
\includegraphics[width=1.0\linewidth]{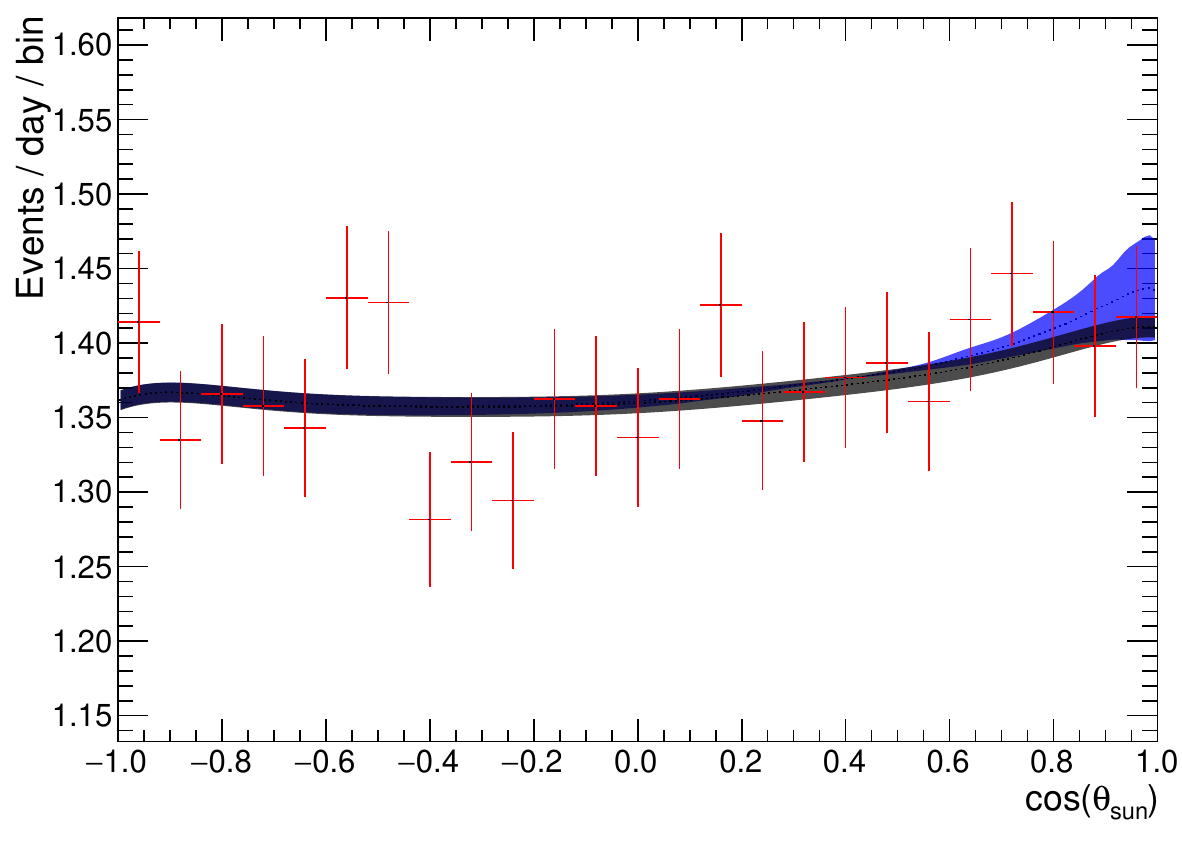}
\caption{\SI{2.99}{MeV}--\SI{3.49}{MeV}, MSG bin 1}
\end{subfigure}
\begin{subfigure}{0.23\linewidth}
\centering
\includegraphics[width=1.0\linewidth]{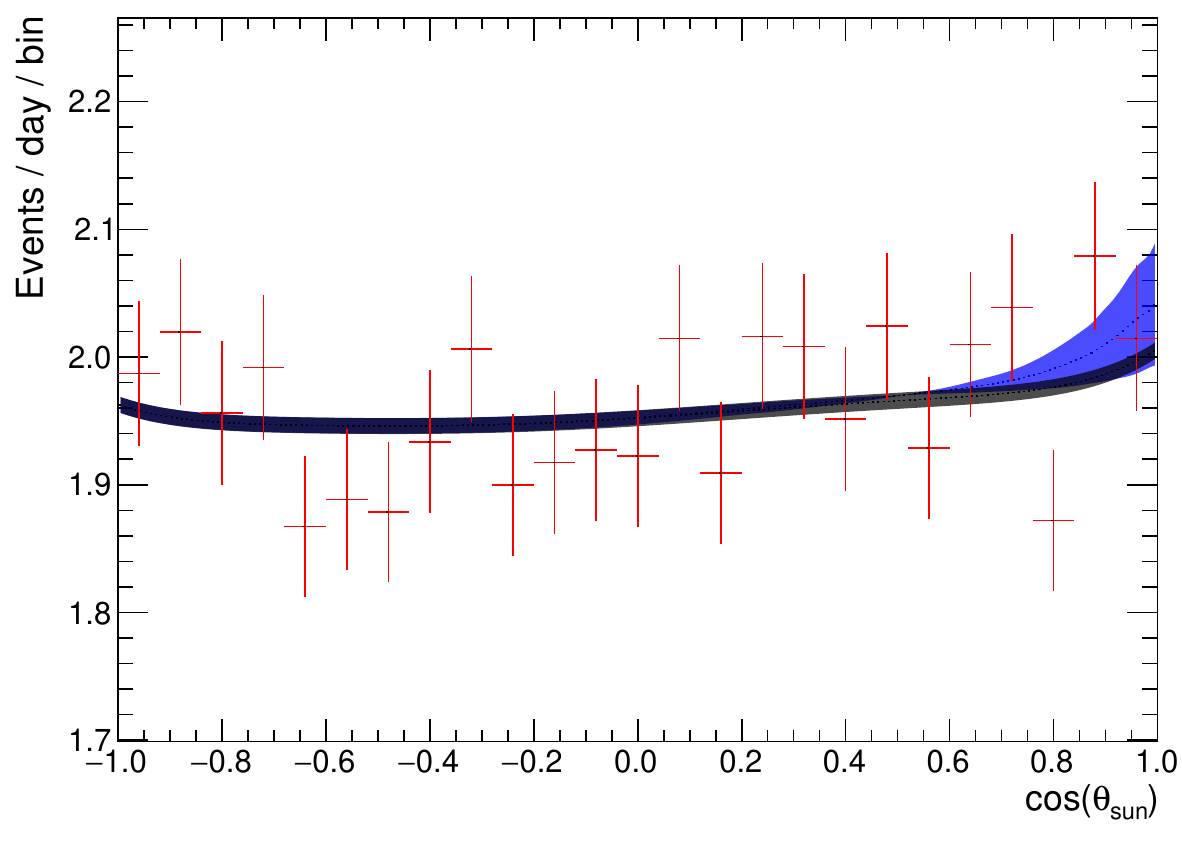}
\caption{\SI{2.99}{MeV}--\SI{3.49}{MeV}, MSG bin 2}
\end{subfigure}
\begin{subfigure}{0.23\linewidth}
\centering
\includegraphics[width=1.0\linewidth]{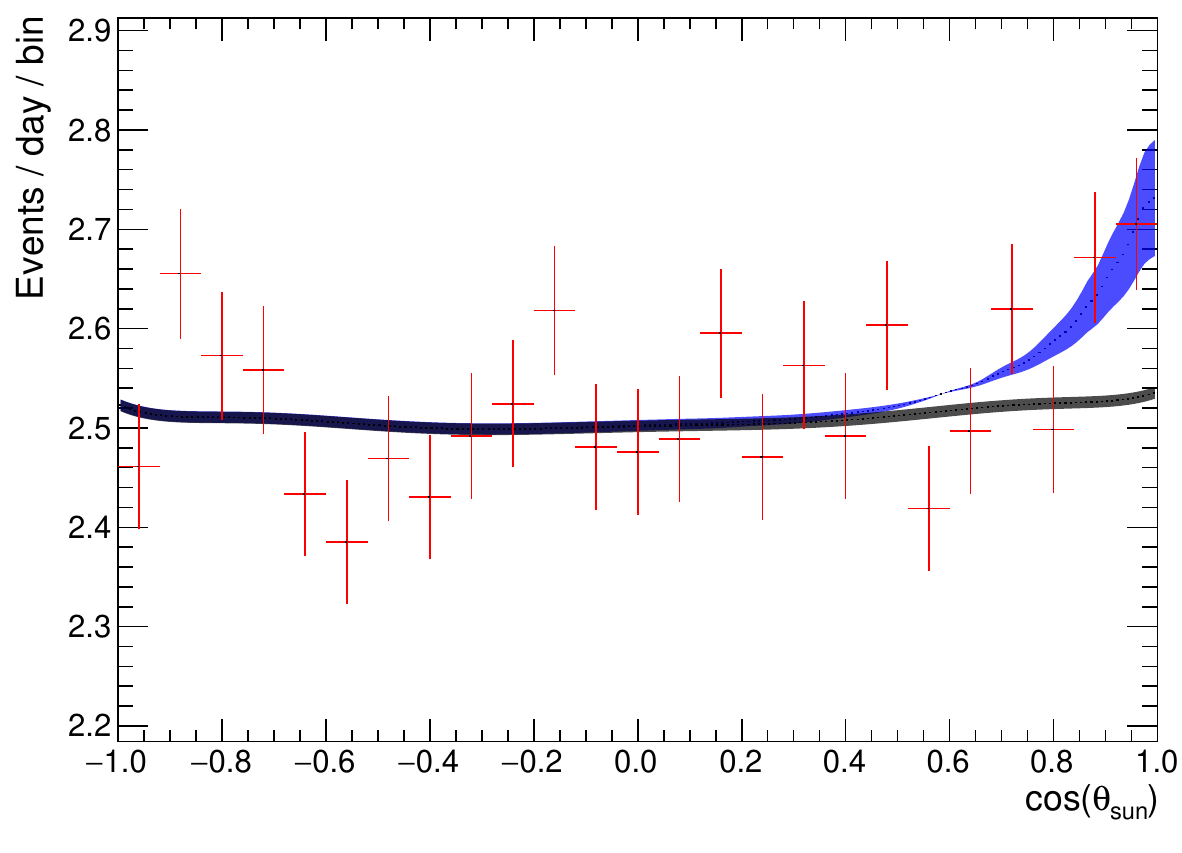}
\caption{\SI{2.99}{MeV}--\SI{3.49}{MeV}, MSG bin 3}
\end{subfigure}
\caption{\label{fig:cossun_wit_msg} Solar angle distributions for the WIT dataset in all MSG bins for data selection (red), calculated background shape (black), and calculated signal shape added to background shape (blue) all with $1\sigma$ statistical error bands.}
\end{figure*}

\section{Signal extraction}\label{sec:solfit}

Following the event selection, the approach in \cite{skiv} is largely followed to measure the number of solar neutrino events. This method uses an extended maximum likelihood function fit on the solar angle distribution based on expected distributions (``shapes'') for simulated signal and background from data. The signal shape is generated using MC with polynomial fits. Two methods are used in \cite{skiv} to generate background shapes. In the ``standard'' method, polynomial fits are applied to the event direction distributions in detector coordinates of the zenith ($\cos{\theta}$) and azimuthal ($\phi$) angles, followed by generating a toy MC and refitting before projecting to the solar angle $(\cos{\theta_{\text{sun}}}$) \cite{ski}. In the ``scramble'' method, the background shape for $\cos{\theta_{\text{sun}}}$ is calculated by filling this solar angle distribution with $\cos{\theta^{ij}_{\text{sun}}} = \hat{d}_i \cdot \hat{s}_j$ for all possible pairs of event direction $\hat{d}_i$ and solar direction $\hat{s}_j$. If the signal-to-background ratio is poor, there will be many more pairs of background events than pairs of signal events. The scramble method is more appropriate than the standard method for a low signal-to-background ratio but becomes less appropriate than the standard method at higher energies with better signal-to-background ratios. The analysis in \cite{skiv} therefore used the standard method for the entire sample while reserving the scramble method as an alternative method for the purpose of calculating the background shape systematic error. The opposite is done here due to the very poor signal-to-background ratio in this very-low-energy range, designating the scramble method as the default and the standard as the alternative for systematic error determination. Both signal and background shapes are converted to polynomial probability density functions prior to the maximum likelihood fit.

The MSG variable (Table \ref{tab:vars}, Index 18) is useful for this signal extraction step as multiple Coulomb scattering leads to worse direction resolution, smearing the solar peak. MSG represents the goodness of a direction fit performed by an alternative direction fitter \cite{skiv_2016}. This fitter reconstructs direction by drawing \ang{42} cones corresponding to the vectors connecting the vertex and each hit PMT within \SI{20}{ns} after time-of-flight subtraction. Clusters of lines of intersection between such cones will form around likely direction vectors, defining candidate directions. The intersection unit vectors assigned to the candidate cluster are added together, and the magnitude of this vector sum is compared to the largest possible: the number of intersections. MSG is the ratio of the two, ranging between 0 and 1, and is intended to quantify the amount of scattering the electron exhibits. Since multiple scattering is dependent on the electron energy, MSG may also be interpreted as an indirect electron energy estimation method.

\begin{figure*}[t]
\centering
\begin{subfigure}{0.23\linewidth}
\centering
\includegraphics[width=1.0\linewidth]{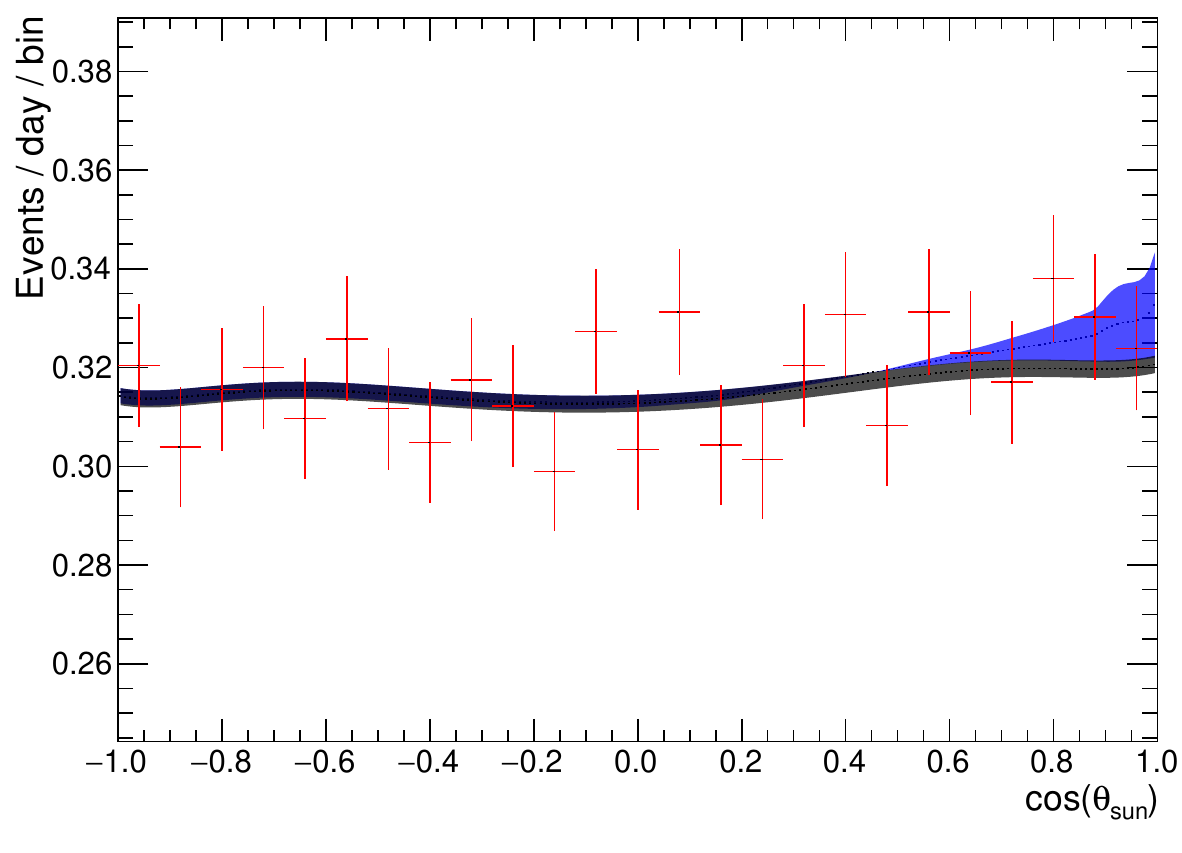}
\caption{34-hit period,\protect\newline MSG bin 1}
\end{subfigure}
\begin{subfigure}{0.23\linewidth}
\centering
\includegraphics[width=1.0\linewidth]{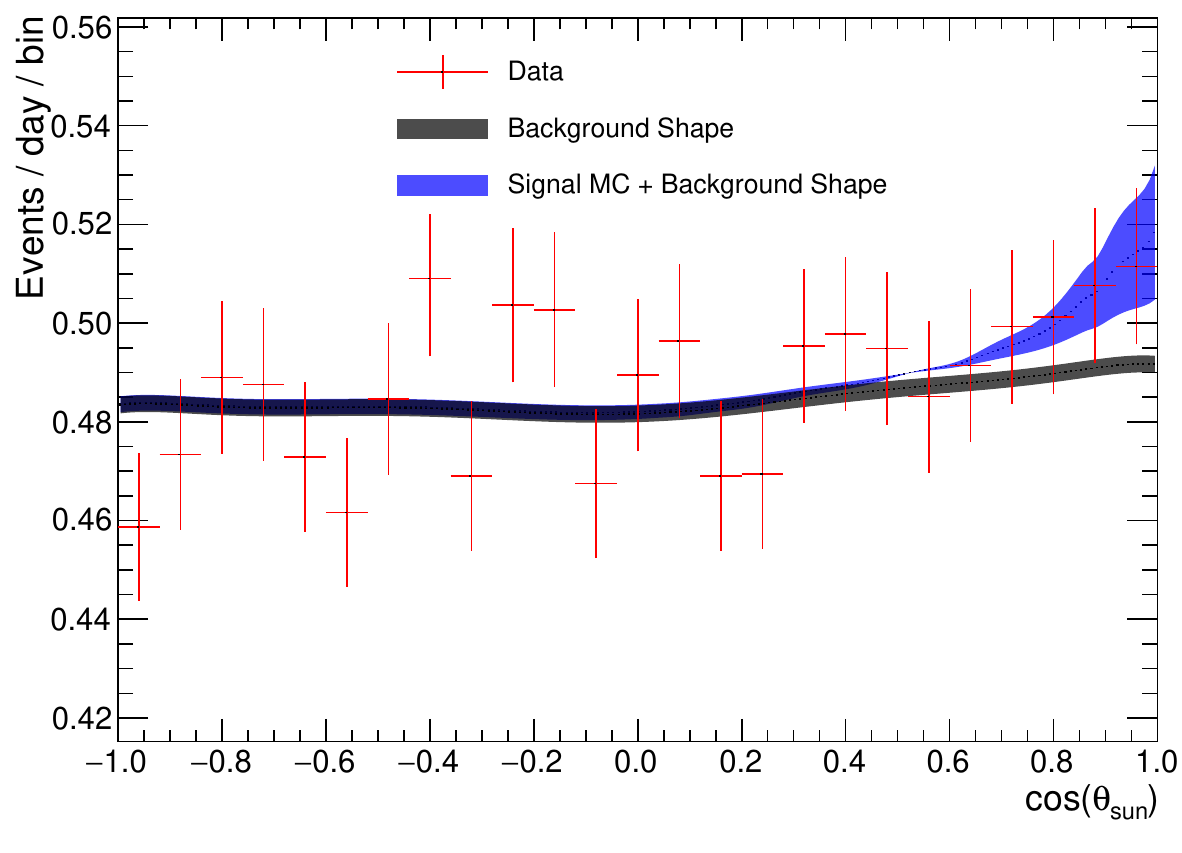}
\caption{34-hit period,\protect\newline MSG bin 2}
\end{subfigure}
\begin{subfigure}{0.23\linewidth}
\centering
\includegraphics[width=1.0\linewidth]{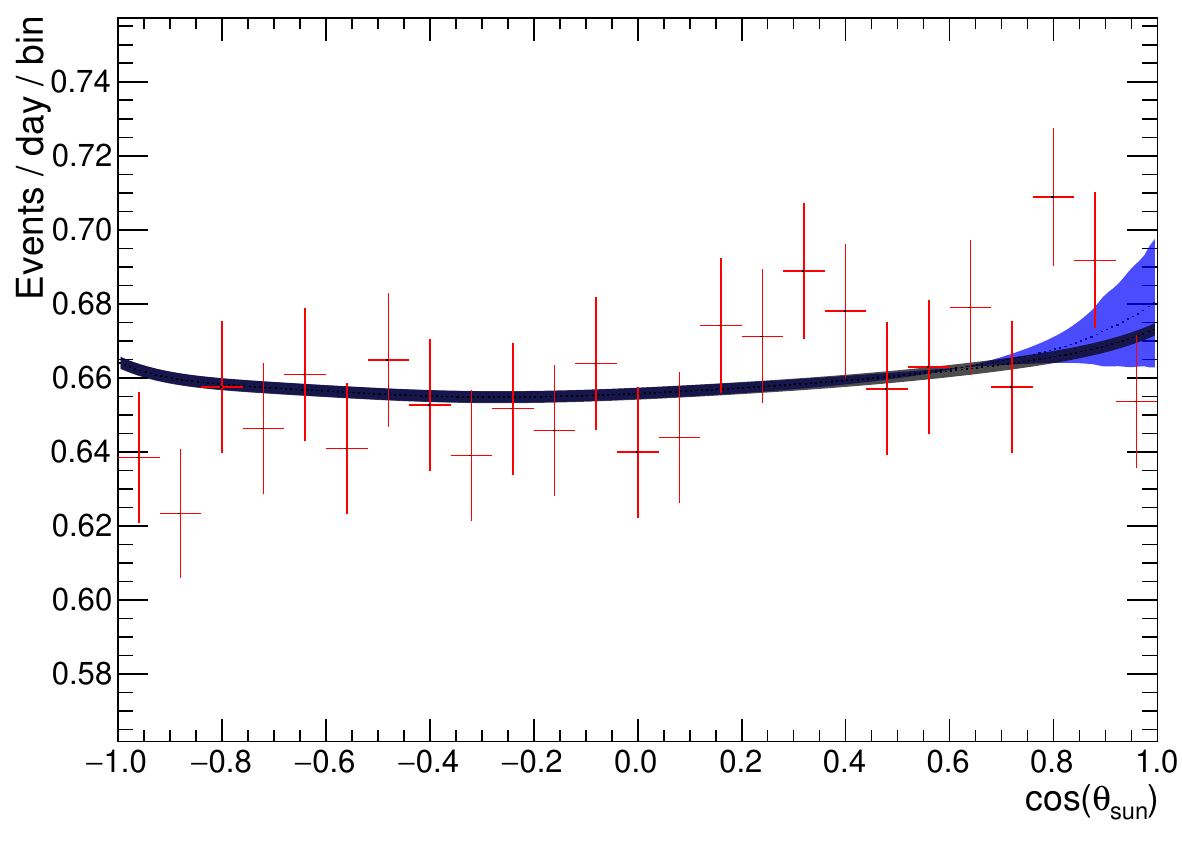}
\caption{\label{fig:cossun_sle_msg3} 34-hit period,\protect\newline MSG bin 3}
\end{subfigure}\\
\begin{subfigure}{0.23\linewidth}
\centering
\includegraphics[width=1.0\linewidth]{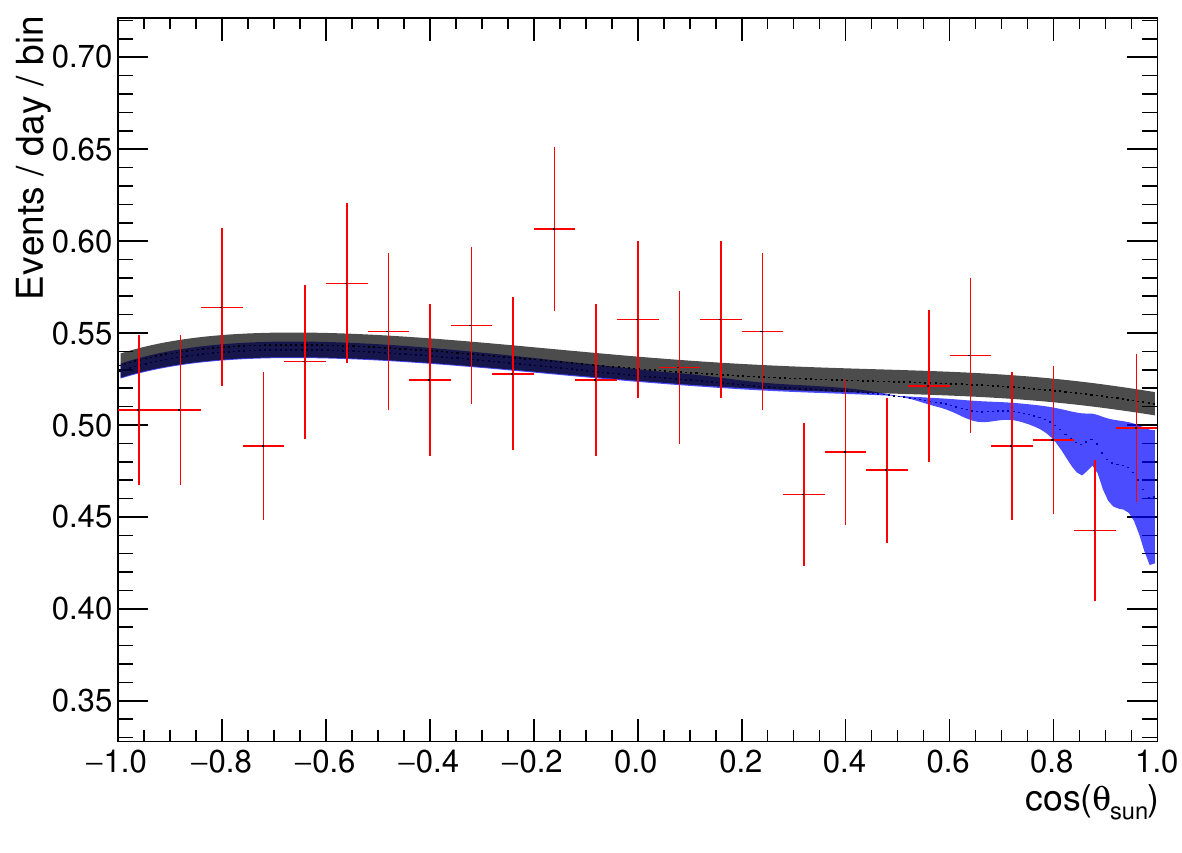}
\caption{31-hit period,\protect\newline MSG bin 1}
\end{subfigure}
\begin{subfigure}{0.23\linewidth}
\centering
\includegraphics[width=1.0\linewidth]{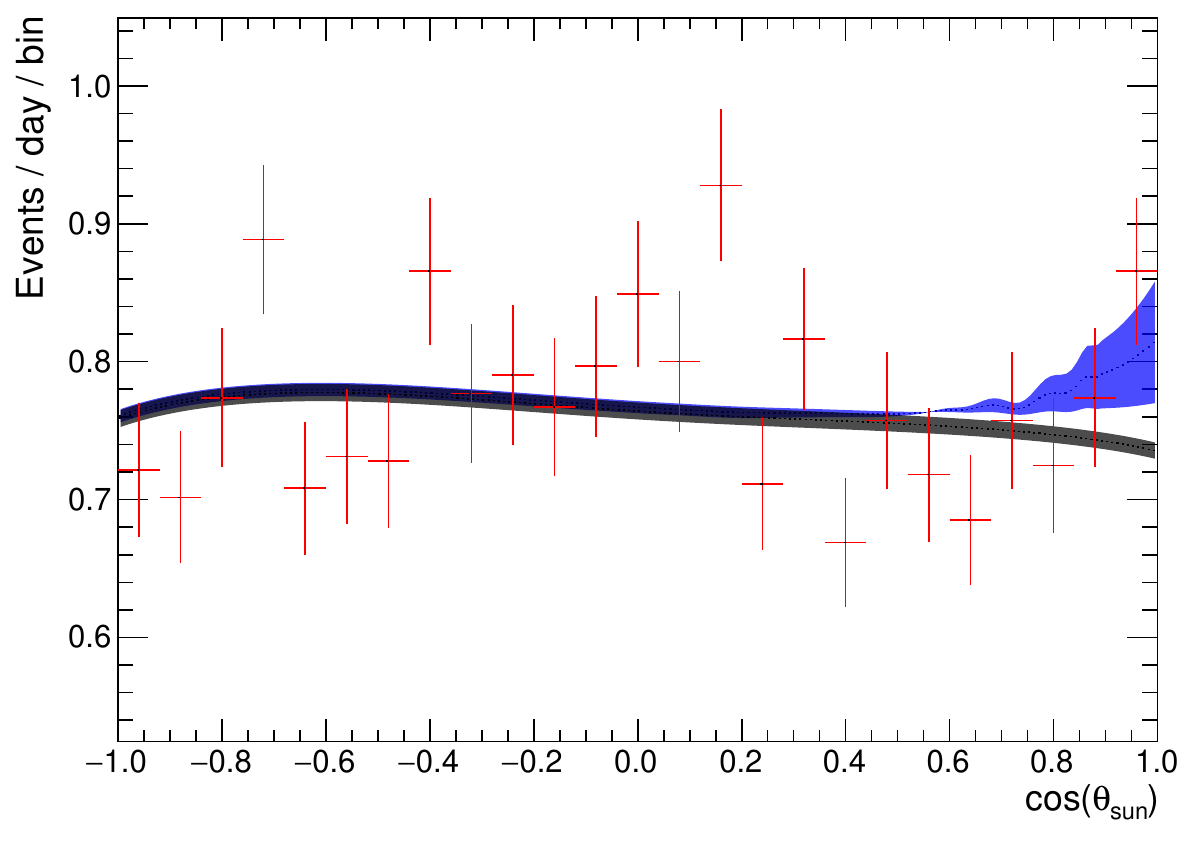}
\caption{31-hit period,\protect\newline MSG bin 2}
\end{subfigure}
\begin{subfigure}{0.23\linewidth}
\centering
\includegraphics[width=1.0\linewidth]{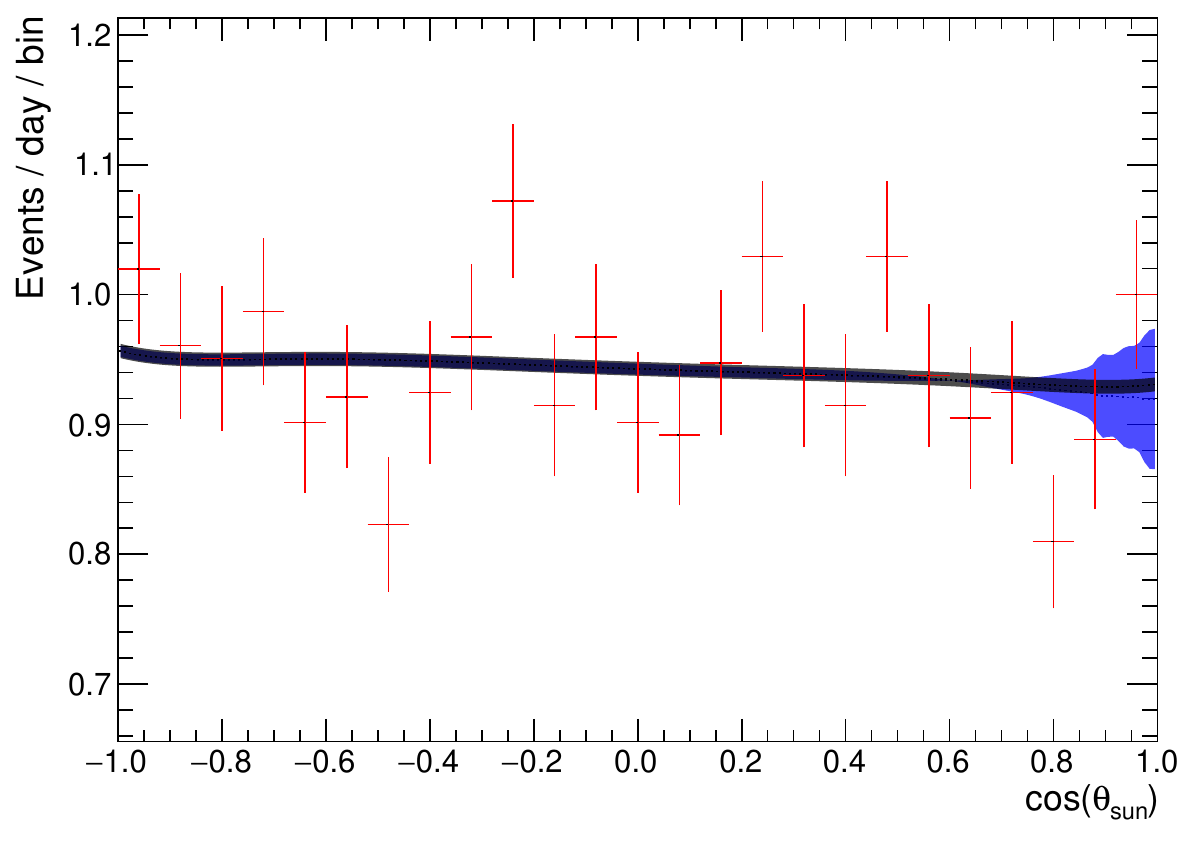}
\caption{31-hit period,\protect\newline MSG bin 3}
\end{subfigure}
\caption{\label{fig:cossun_sle_msg} Solar angle distributions for the SLE dataset in all MSG bins separated by trigger threshold period. The
definition of colors is the same as in Fig. \ref{fig:cossun_wit_msg}.}
\end{figure*}

\begin{figure}[t]
\centering
\begin{subfigure}{0.49\linewidth}
\centering
\includegraphics[width=1.0\linewidth]{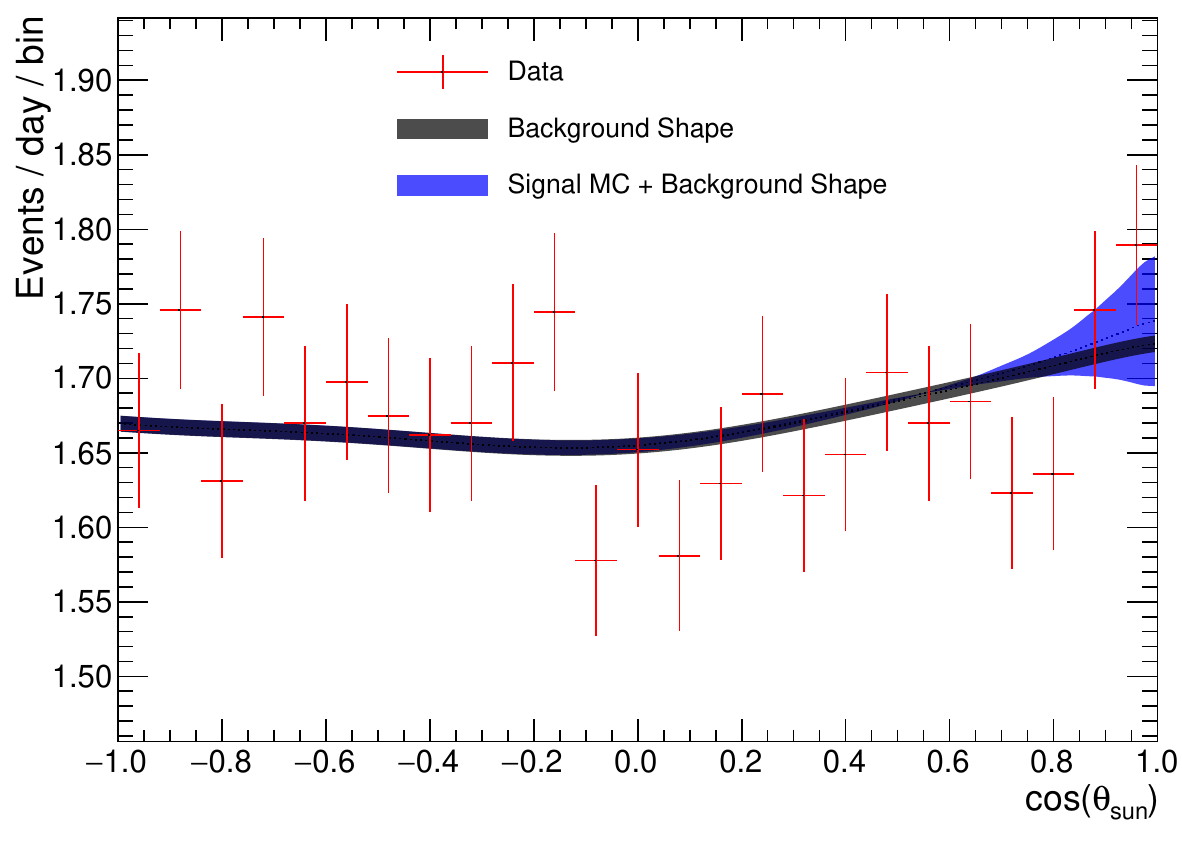}
\caption{\SI{2.49}{MeV}--\SI{2.99}{MeV}}
\end{subfigure}
\begin{subfigure}{0.49\linewidth}
\centering
\includegraphics[width=1.0\linewidth]{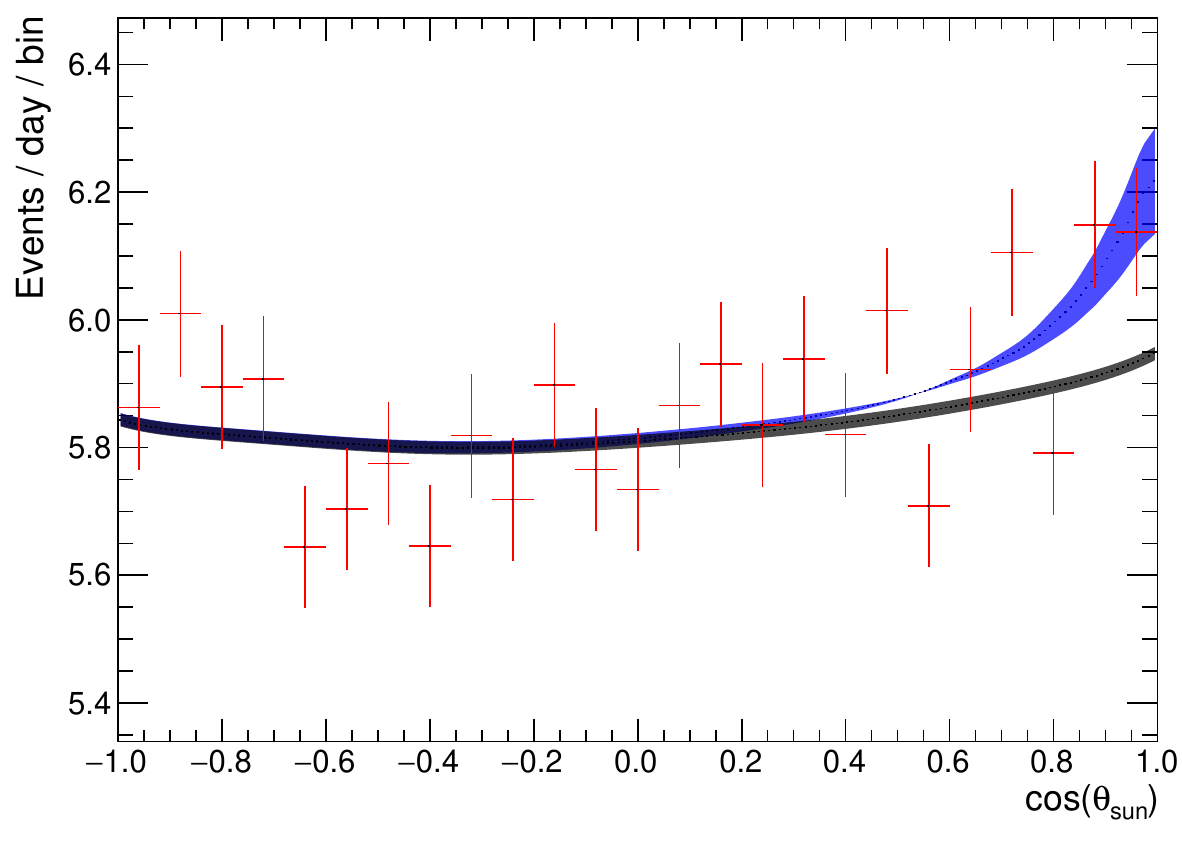}
\caption{\SI{2.99}{MeV}--\SI{3.49}{MeV}}
\end{subfigure}
\caption{\label{fig:cossun_wit} Solar angle distributions aggregated over MSG bins for the WIT dataset. The
definition of colors is the same as in Fig. \ref{fig:cossun_wit_msg}.}
\end{figure}

\begin{figure}[t]
\centering
\begin{subfigure}{0.49\linewidth}
\includegraphics[width=1.0\linewidth]{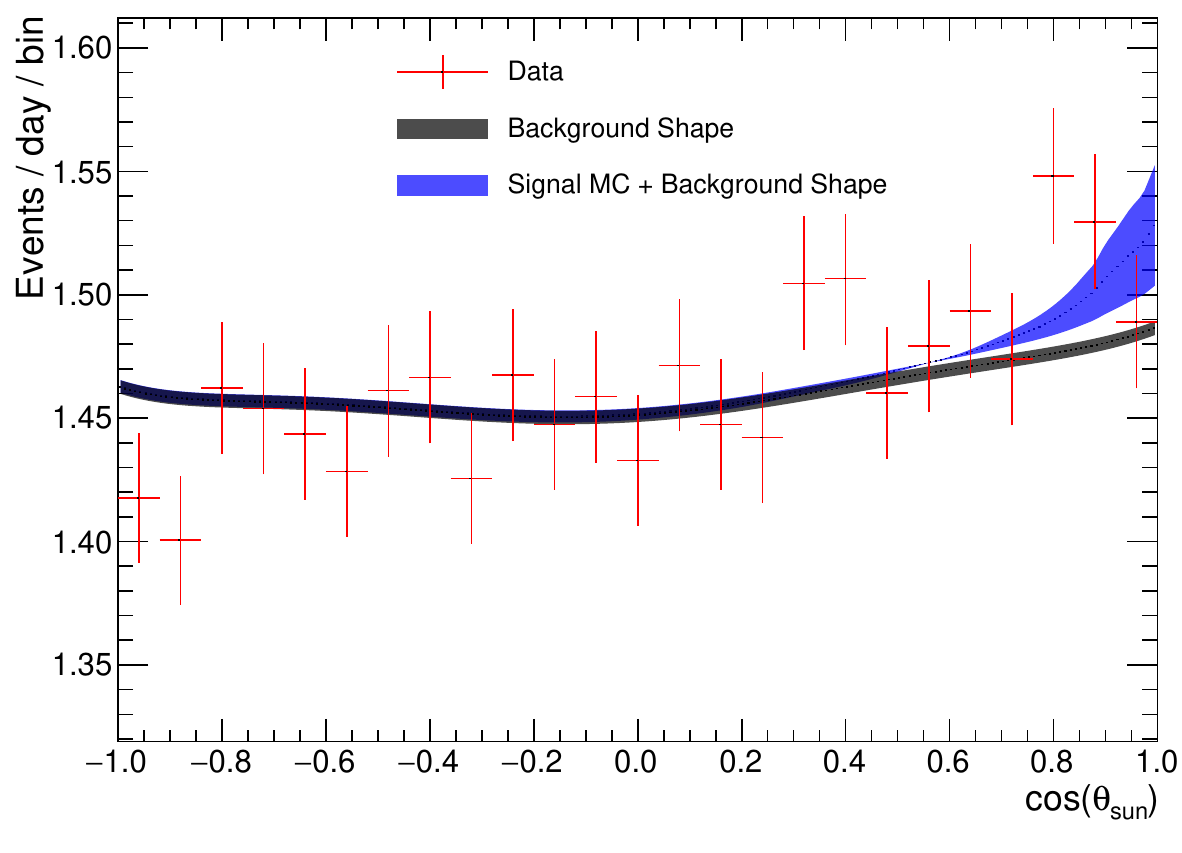}
\caption{34-hit period}
\centering
\end{subfigure}
\begin{subfigure}{0.49\linewidth}
\centering
\includegraphics[width=1.0\linewidth]{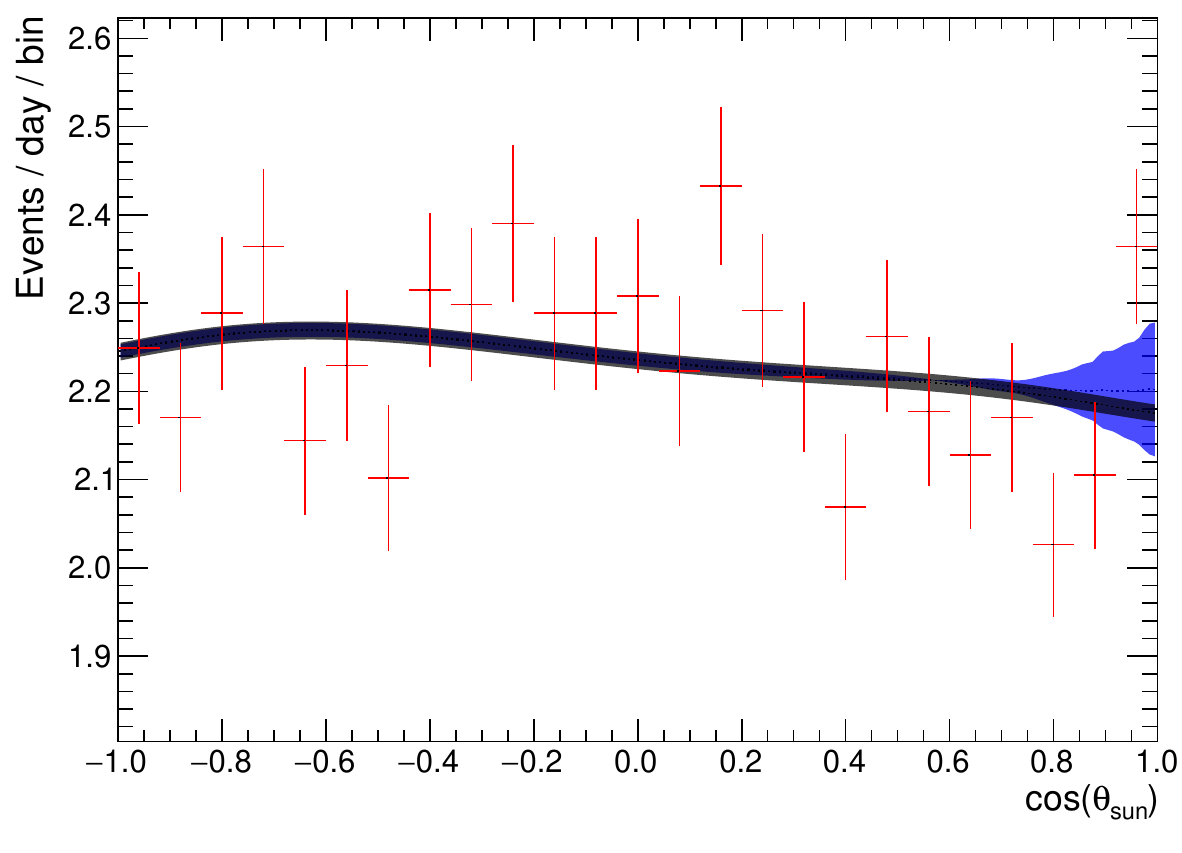}
\caption{31-hit period}
\end{subfigure}
\caption{\label{fig:cossun_sle} Solar angle distributions aggregated over MSG bins for the SLE datasets in \SI{2.99}{MeV}--\SI{3.49}{MeV} separated by trigger threshold period. The
definition of colors is the same as in Fig. \ref{fig:cossun_wit_msg}.}
\end{figure}

The binning of \cite{skiv} is also followed and is simply extended to this energy region of interest: \SI{0.5}{MeV} bins and bins of 0--0.35, 0.35--0.45, and 0.45--1 in MSG, chosen to evenly split the signal content before selection cuts. The fit is applied simultaneously over each MSG bin, and the solar angle distributions with the signal and background shapes in these bins are shown in Figs. \ref{fig:cossun_wit_msg} and \ref{fig:cossun_sle_msg}. In order to demonstrate the presence of solar neutrinos, the signal and background shapes are aggregated over the three MSG bins by weighting each bin according to its associated statistical error of a solar signal observation. Figures \ref{fig:cossun_wit} and \ref{fig:cossun_sle} show these aggregated solar angle distributions.

\section{Systematic uncertainties}

The systematic error calculation methods relating to aspects of the analysis that are unchanged compared to previous analyses are also unchanged. Methods for calculating systematic errors that had to be modified for this analysis are explained in this section.

\subsection{Background shape}\label{sec:systs_backshape}
The uncertainty in the background shape calculation is the largest source of systematic uncertainty. While this error is 2.7\% in the lowest \SI{3.49}{MeV}--\SI{3.99}{MeV} energy bin of \cite{skiv}, the signal-to-background ratio is considerably worse in the next lowest energy bin here. Similarly small fluctuations in the background shape in the region of the solar peak can result in large differences in the fit number of signal events when the solar peak is also small.

\begin{figure*}[t]
\centering
\begin{subfigure}{0.23\linewidth}
\centering
\includegraphics[width=1.0\linewidth]{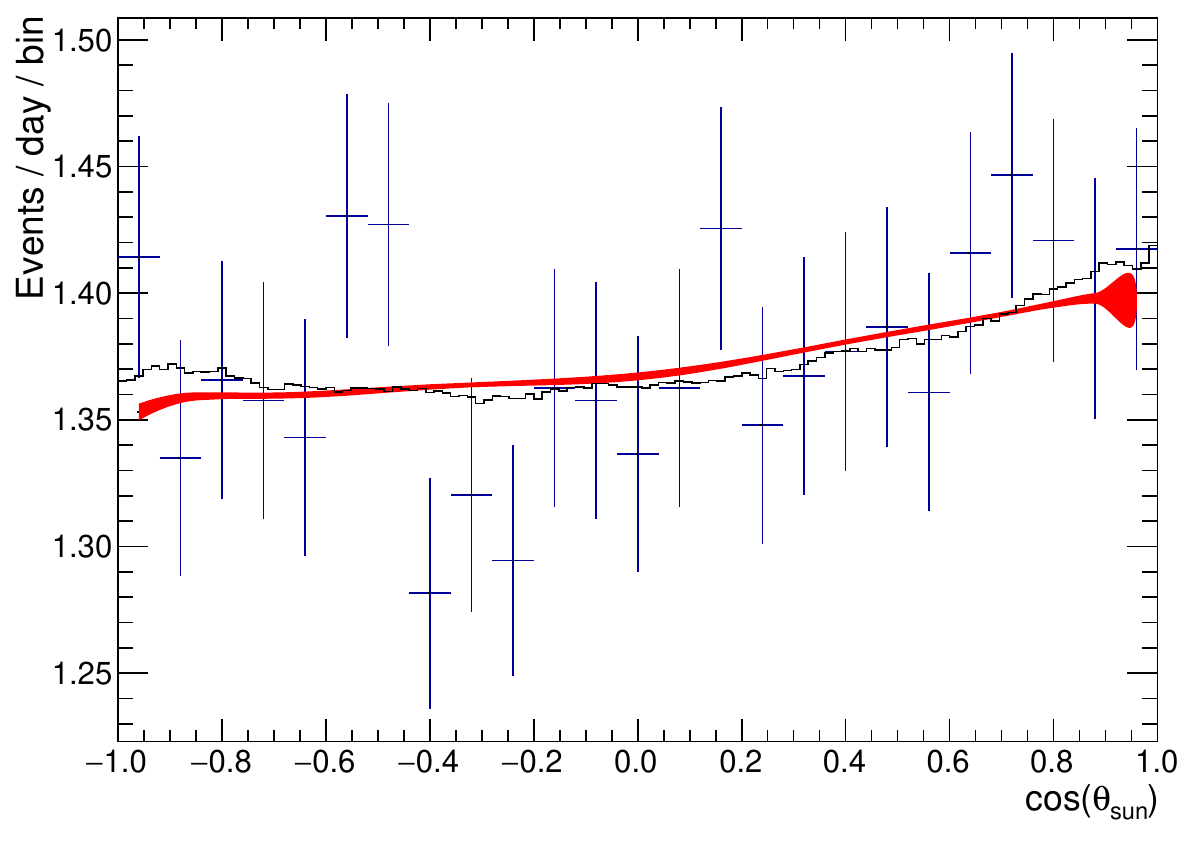}
\caption{WIT, MSG bin 1}
\end{subfigure}
\begin{subfigure}{0.23\linewidth}
\centering
\includegraphics[width=1.0\linewidth]{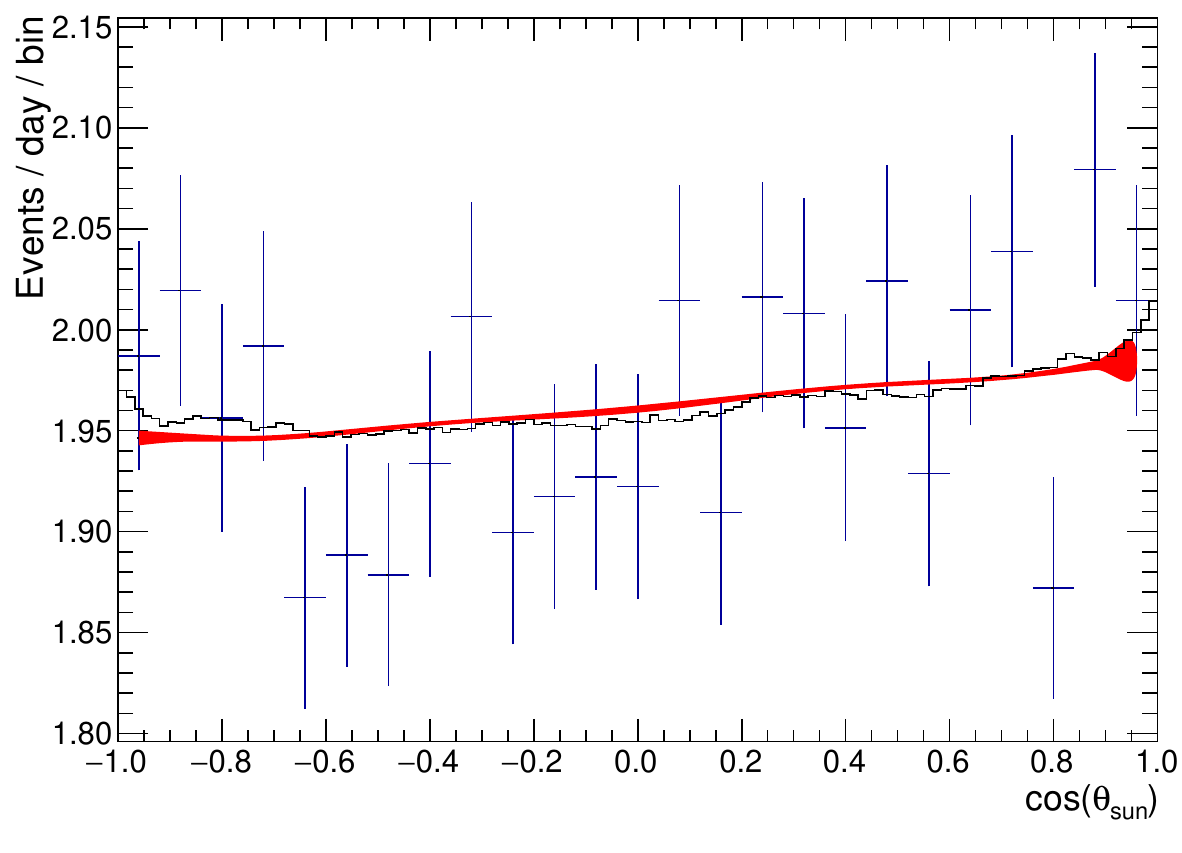}
\caption{WIT, MSG bin 2}
\end{subfigure}
\begin{subfigure}{0.23\linewidth}
\centering
\includegraphics[width=1.0\linewidth]{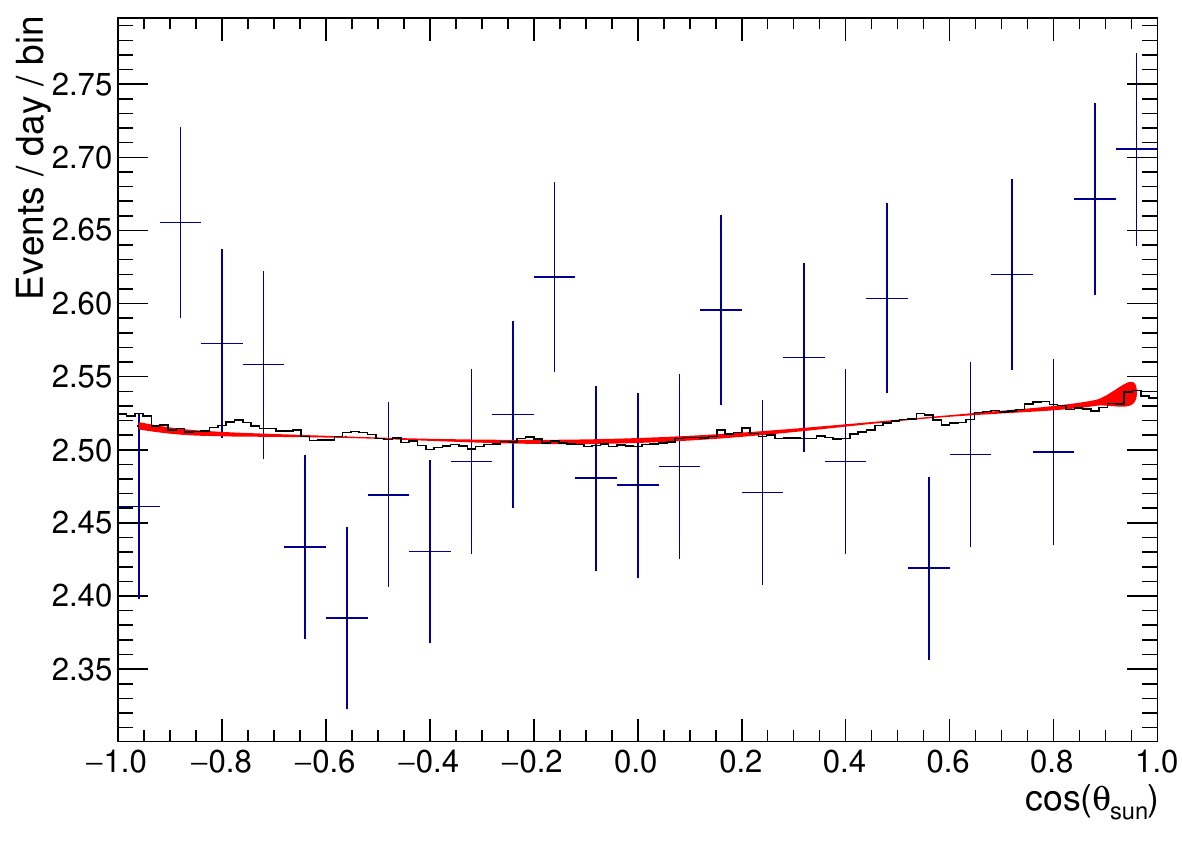}
\caption{WIT, MSG bin 3}
\end{subfigure}\\
\begin{subfigure}{0.23\linewidth}
\centering
\includegraphics[width=1.0\linewidth]{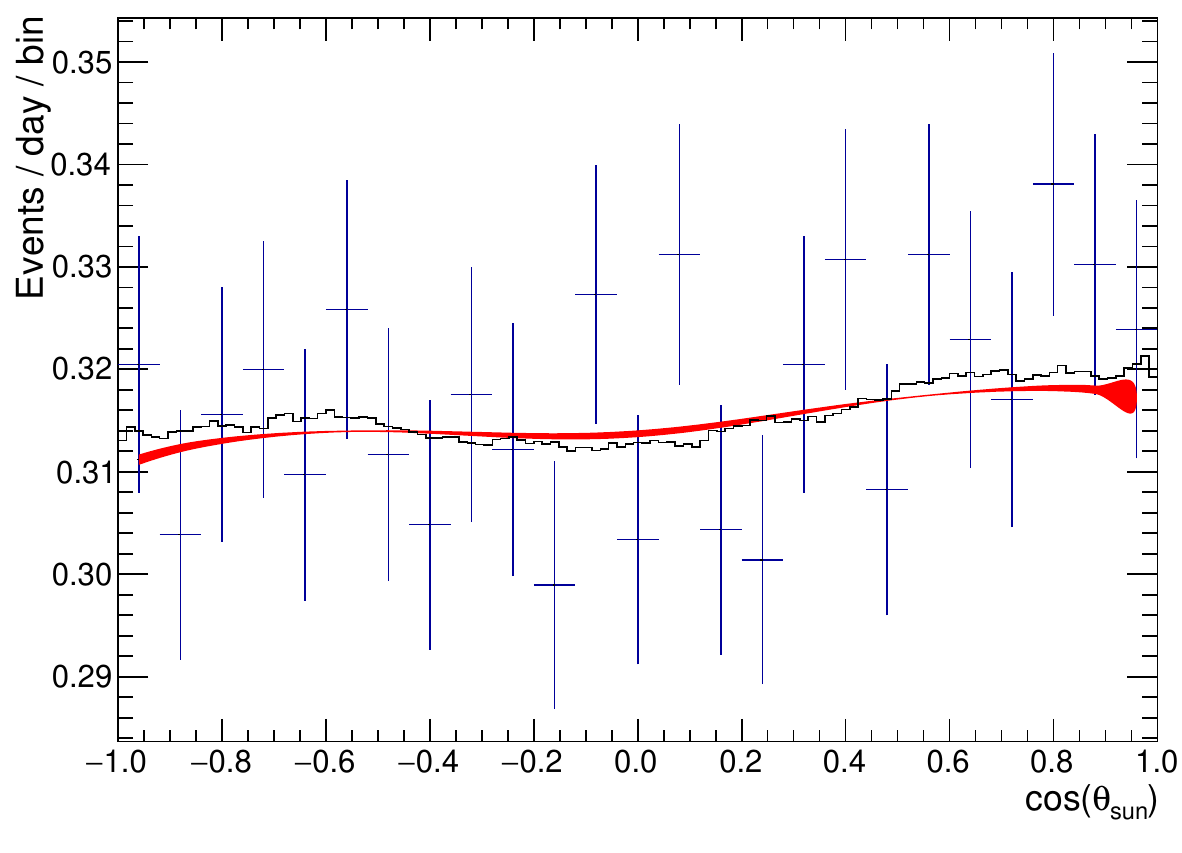}
\caption{SLE 34-hit period, \protect\newline MSG bin 1}
\end{subfigure}
\begin{subfigure}{0.23\linewidth}
\centering
\includegraphics[width=1.0\linewidth]{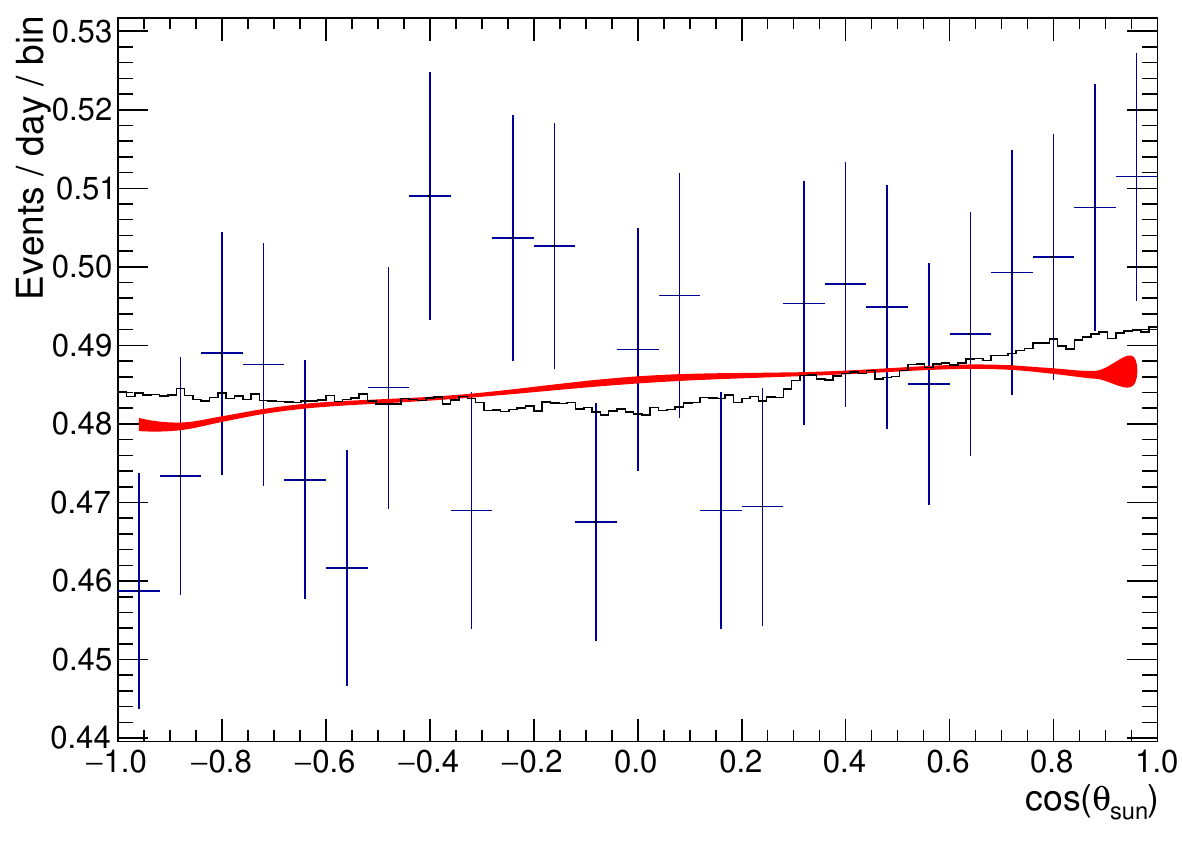}
\caption{SLE 34-hit period, \protect\newline MSG bin 2}
\end{subfigure}
\begin{subfigure}{0.23\linewidth}
\centering
\includegraphics[width=1.0\linewidth]{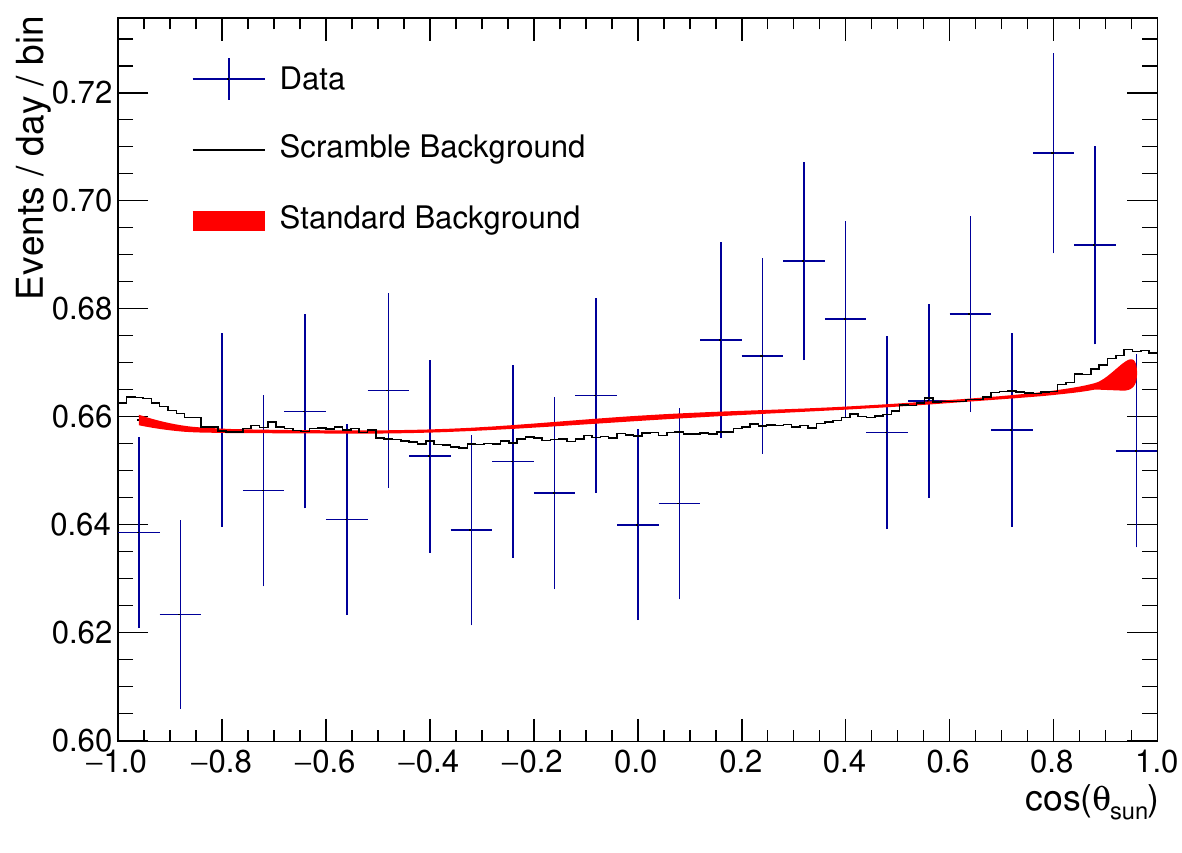}
\caption{SLE 34-hit period, \protect\newline MSG bin 3}
\end{subfigure}
\caption{\label{fig:backshape_err} Comparison of background shape calculation methods for systematic error calculation in each MSG bin for the WIT and SLE 34-hit period in \SI{2.99}{MeV}--\SI{3.49}{MeV}. The blue points show the data, the black line shows the scramble method, and the red band shows the standard method.}
\end{figure*}

We compare the resulting number of signal events when using the two background shape methods, $N_{\text{scramble}}$ and $N_{\text{standard}}$, and calculate the error as
$$\frac{N_{\text{scramble}} - N_{\text{standard}}}{N_{\text{scramble}}} (\times100\%),$$
taking the scramble method as nominal (see Sec. \ref{sec:solfit}).
Figure \ref{fig:backshape_err} shows the comparison of the scramble and standard shapes for all bins. The background shape errors in \SI{2.99}{MeV}--\SI{3.49}{MeV} for the full MSG 0--1 range were 11.2\% for WIT and 41.8\% for 34-hit SLE. We do not calculate an error for the 31-hit period (no WIT) or the WIT \SI{2.49}{MeV}--\SI{2.99}{MeV} bin, as no significant signal was observed (see Sec. \ref{sec:flux}).

\subsection{Trigger efficiency}
The calculation of the trigger efficiency error in SK-IV is based on calibration data taken in 2015 with a Ni gamma ray source \cite{calib}. The trigger efficiency of Ni events at various points is compared between data and MC, and the efficiency error at each point is $e_i$. The BDTs used for the 34-hit and 31-hit SLE datasets were applied to all Ni data and MC from the 2015 calibration runs, including those outside of the \SI{8.5}{kton} tight fiducial volume region. The same BDT cut used during solar event selection was then applied to data, giving the number of surviving events $B_i$, adjusting for the live time of each Ni run. We determine the overall trigger efficiency systematic error as $\frac{\sum{e_i B_i}}{\sum{B_i}}$. This error is 16.2\% for the 34-hit threshold and 7.7\% for the 31-hit threshold.

For the WIT dataset, the trigger efficiency error should be small because the WIT pretrigger efficiency is near 100\%. The efficiency of the online WIT cuts from Fig. \ref{fig:wit_eff} in the relevant bins is given in Table \ref{tab:trigeff_wit}.

\begin{table}[b]
\caption{\label{tab:trigeff_wit}Trigger efficiency of the online WIT cuts for solar MC.}
\begin{ruledtabular}
\begin{tabular}{lccc}
$E_{\text{kin}}$ Range (MeV) & 3.49--3.99 & 2.99--3.49 & 2.49--2.99\\
\colrule
Efficiency & 94.3\% & 91.4\% & 83.0\% \\
\end{tabular}
\end{ruledtabular}
\end{table}

While the systematic error of cuts on online reconstructed quantities is expected to be lower than that of a simple hit coincidence trigger, and such systematics should be covered by shifts to the relevant variables in Sec. \ref{sec:systs_bdt}, there is no extensive set of calibration data available during the time WIT was online during SK-IV, so the method applied for SLE data cannot be used for WIT to verify these expectations. A very conservative trigger efficiency error is therefore determined by assigning the actual trigger inefficiency as the systematic error. This inefficiency is 8.6\% in \SI{2.99}{MeV}--\SI{3.49}{MeV} (Table \ref{tab:trigeff_wit}).

\subsection{BDT input variables}\label{sec:systs_bdt}
To assign systematic errors for the reconstruction variables input to the BDT arising from discrepancies between solar signal MC and data, these variables are shifted by an amount motivated by calibration data, and the change in efficiency of the BDT output cut is assigned as the systematic error. The discrepancy in a particular variable between LINAC \cite{linac} or Ni \cite{calib} calibration data and MC determines the amount of the shift, which is energy dependent. Since BDTs consider all input variables simultaneously, the BDTs were reevaluated after shifting variables one at a time for those variables that are not correlated with any others to first order. However, the shifts were propagated for highly related variables. The only input variable with appreciable systematic error calculated in this way is the reconstruction goodness $g_t^2-g_p^2$ due to its high separation power for signal vs background \cite{skiv_2016}.

\subsection{Error summary}
Table \ref{tab:systs} shows a summary of the systematic errors for the \SI{2.99}{MeV}--\SI{3.49}{MeV} WIT and SLE 34-hit datasets in which a solar signal was observed in Sec. \ref{sec:results}. Other sources of error that were evaluated but were less significant include signal extraction, $^8$B spectrum \cite{b8_spec}, energy resolution, and the rest of the BDT input variables. The ``total'' systematic error is quoted for the interaction rate measurements (Sec. \ref{sec:flux}). The ``total uncorrelated'' errors along with the energy-correlated errors, as calculated in \cite{skiv}, extended down to \SI{2.99}{MeV} are used for the energy spectrum fit (Sec. \ref{sec:spectrum}).

\begin{table}[h]
\caption{\label{tab:systs}Systematic error summary with significant error sources. ``Total'' includes all errors added in quadrature, and ``total uncorrelated'' excludes the energy scale error.}
\begin{ruledtabular}
\begin{tabular}{lcc}
Error Source              & WIT & 34-hit SLE\\
\colrule
Background shape          &  11\% &  42\%\\
Trigger efficiency        &  8.6\% &  16\%\\
$g_t^2-g_p^2$ fit goodness& 2.1\% & 2.1\%\\
Energy scale              & 4.5\% & 10\%\\
\colrule
Total uncorrelated        & 14\% & 45\%\\
Total                     & 15\% & 46\%\\
\end{tabular}
\end{ruledtabular}
\end{table}

\section{Results}\label{sec:results}

\subsection{Interaction rate measurements}\label{sec:flux}

\renewcommand{\arraystretch}{1.3}
\begin{table*}[t]
    \caption{Expected number of unoscillated MC signal events, fit number of signal events, and their ratio in each energy and MSG bin. Measurements are given as measurement $\pm$ statistical $\pm$ systematic error. Systematic errors are only calculated for the 0--1 MSG bins, and only if a signal is observed. When aggregating across all MSG bins, bins are weighted by their statistical error. When aggregating across the datasets, datasets are weighted by their total error.}\label{tab:fluxes}
	\begin{ruledtabular}
	    \begin{tabular}{cccccc}
			Dataset & \textit{E}\textsubscript{\text{kin}} bin (MeV)	& MSG bin     & Expected (unoscillated) & Observed & Observed/Expected\\
			\hline
			\multirow{8}{*}{WIT}& 2.49--2.99	& 0--0.35 & 41 & $-45^{+46}_{-48}$ & $-1.1^{+1.1}_{-1.2}$\\
			&2.49--2.99	& 0.35--0.45 & 71 & $-13^{+48}_{-50}$ & $-0.18^{+0.68}_{-0.70}$\\
			&2.49--2.99	& 0.45--1 & 115 & $49^{+54}_{-53}$ & $0.42^{+0.47}_{-0.46}$\\
			\cline{2-6}
			&2.99--3.49	& 0--0.35 & 247 &  $63^{+104}_{-102}$ & $0.25^{+0.42}_{-0.41}$\\
			&2.99--3.49	& 0.35--0.45 & 362  & $64^{+97}_{-96}$ & $0.17^{+0.27}_{-0.26}$\\
			&2.99--3.49	& 0.45--1 & 467 &  $276^{+92}_{-90}$ & $0.59^{+0.20}_{-0.19}$\\
			\cline{2-6}
			&2.49--2.99	& 0--1 & 227  & $25^{+83}_{-81}$ & $0.11^{+0.37}_{-0.36}$\\
			&2.99--3.49	& 0--1 & 1077 &  $457^{+159}_{-157}\pm69$ & $0.425^{+0.148}_{-0.146}\pm0.064$\\
			\hline
			\multirow{4}{*}{SLE 34-hit}&2.99--3.49	& 0--0.35 & 230 &  $85^{+90}_{-88}$ & $0.37^{+0.40}_{-0.38}$\\
			&2.99--3.49	& 0.35--0.45 & 356 & $151^{+88}_{-86}$ & $0.42^{+0.25}_{-0.24}$\\
			&2.99--3.49	& 0.45--1 & 506 &  $30^{+83}_{-82}$ & $0.06^{+0.16}_{-0.16}$\\
			\cline{2-6}
            &2.99--3.49	& 0--1 & 1092 &  $213^{+141}_{-140}\pm98$ & $0.196^{+0.130}_{-0.128}\pm0.090$\\
            \hline
            \multirow{4}{*}{SLE 31-hit}&2.99--3.49	& 0--0.35 & 70 & $-61^{+48}_{-51}$ & $-0.86^{+0.68}_{-0.72}$\\
			&2.99--3.49	& 0.35--0.45 & 99  & $73^{+48}_{-46}$ & $0.73^{+0.49}_{-0.47}$\\
			&2.99--3.49	& 0.45--1 & 121 &  $-8^{+41}_{-43}$ & $-0.07^{+0.34}_{-0.35}$\\
			\cline{2-6}
            &2.99--3.49	& 0--1 & 291 &  $23^{+77}_{-74}$ & $0.08^{+0.26}_{-0.26}$\\
            \hline
            WIT + 34-hit&2.99--3.49	& 0--1 & 2169 &  $667^{+195}_{-198}\pm143$ & $0.307^{+0.091}_{-0.090}\pm0.066$\\
		\end{tabular}
    \end{ruledtabular}
\end{table*}

Table \ref{tab:fluxes} shows the results for each energy and MSG bin for all datasets as well as the results aggregated over MSG bins and combined over multiple datasets. The number of observed signal events is compared to the expected number of unoscillated signal events calculated using the same approach as \cite{skiv} based on the neutral current results from SNO \cite{sno_nc}.

No statistically significant indication of a signal is found in the {2.49}{MeV}--\SI{2.99}{MeV} bin due to the large amount of radioactive background and in the 31-hit period \SI{2.99}{MeV}--\SI{3.49}{MeV} bin since there is little live time left after removing runs that are available in WIT. In the \SI{2.99}{MeV}--\SI{3.49}{MeV} bin of the other datasets, there is a $2.66\sigma$ significance for a solar signal in WIT and $1.25\sigma$ in the 34-hit period after accounting for systematic errors. Together, these results give a $2.76\sigma$ indication for a solar signal in this energy bin for SK-IV. The event rate measured in this electron kinetic energy bin corresponds to a solar neutrino flux of $(1.61^{+0.48}_{-0.47} \, \rm{(stat.)} \pm 0.35 \, \rm{(syst.)}) \times 10^{6}~\mathrm{cm^{-2}\,s^{-1}}$ for the entire $^8$B neutrino energy range.

\begin{figure}[t]
\centering
\includegraphics[width=0.95\linewidth]{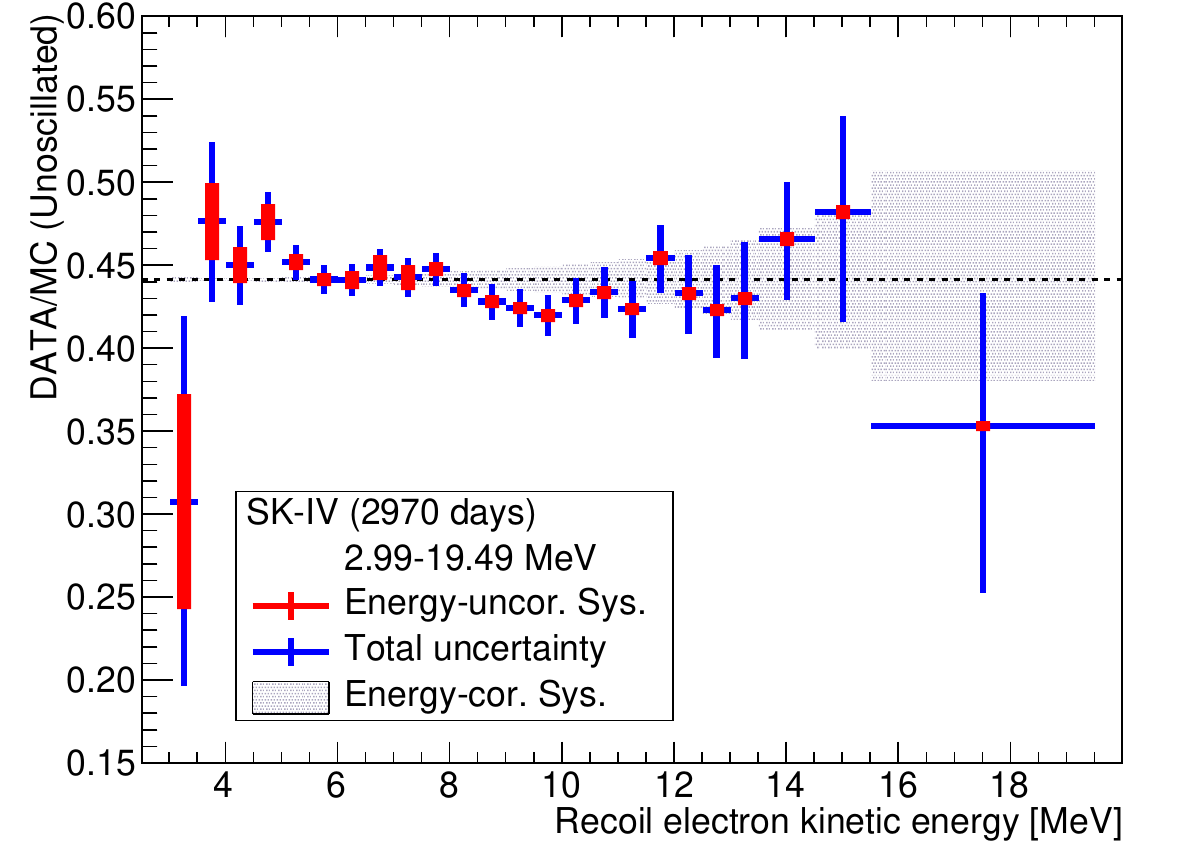}
\caption{\label{fig:spec}The measured electron kinetic energy spectrum in SK-IV with the addition of the lowest energy bin. The blue points and bars show the observed rate divided by the expected event rate assuming no neutrino oscillation, with statistical and energy-uncorrelated uncertainties. The red bars and gray bands show the energy-uncorrelated and energy-correlated systematic uncertainties. The dashed black line shows the average ratio.}
\end{figure}

\begin{figure}[t]
\centering
\includegraphics[width=0.95\linewidth]{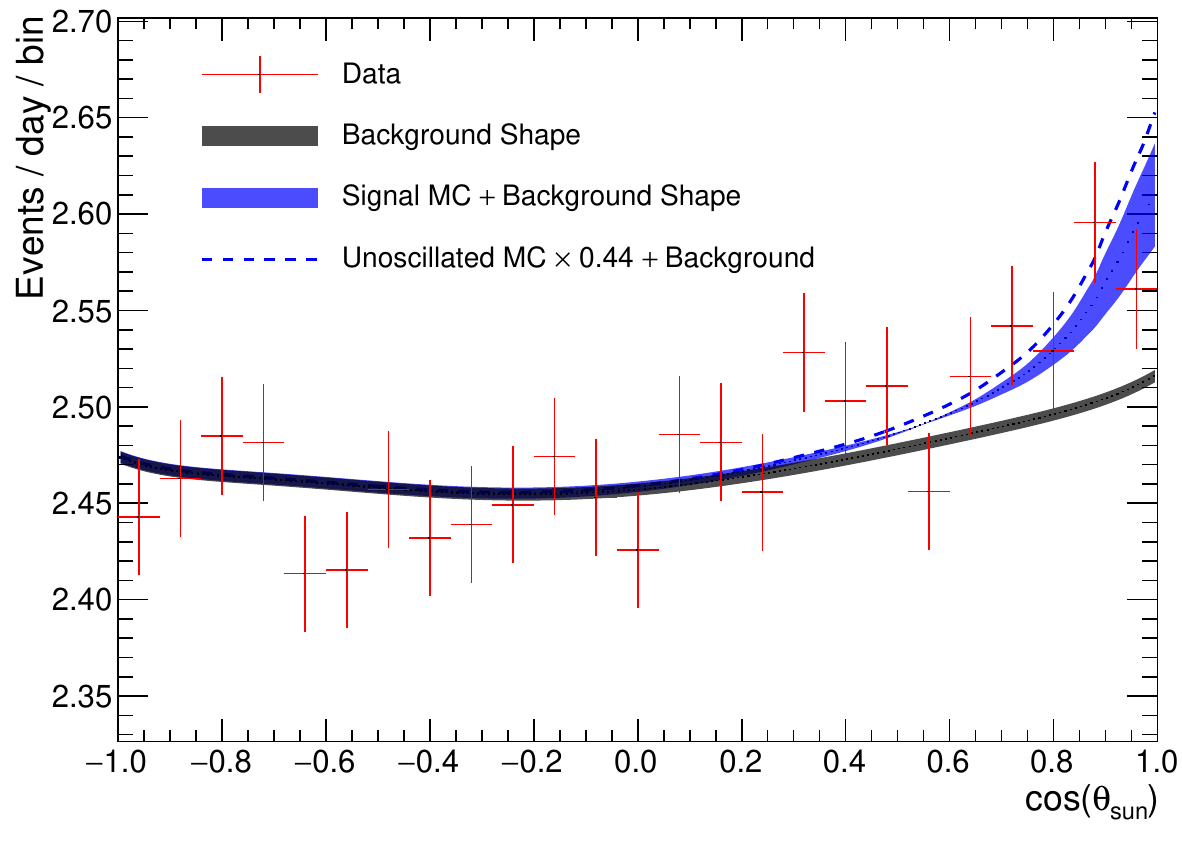}
\caption{\label{fig:cossun_total} Solar angle distribution in \SI{2.99}{MeV}--\SI{3.49}{MeV} aggregated over the WIT and SLE 34-hit datasets. The
definition of colors is the same as in Fig. \ref{fig:cossun_wit_msg}. The additional blue dashed line indicates the signal shape if the average interaction rate in \SI{3.49}{MeV}--\SI{19.49}{MeV} measured in \cite{skiv} is assumed.}
\end{figure}

In comparing the observed number of signal events to the expected unoscillated number, Fig. \ref{fig:spec} shows the measured ratio compared to those of higher-energy bins. The combined ratio in this energy bin has a slight deficit compared to both the 0.44 ratio expected for an energy-independent survival probability ($1.20\sigma$) and the previous 0.47 best-fit prediction for the case of an upturn ($1.47\sigma$) \cite{skiv}. Figure \ref{fig:cossun_total} shows the combined solar angle distribution along with the distribution implied in the 0.44 ratio case. Other datasets and MSG bins generally have similar or better agreement with the expected ratio. However, there is a notable exception in the highest MSG bin of the 34-hit dataset [Table \ref{tab:fluxes} and Fig. \ref{fig:cossun_sle_msg}(c)]. The lower two MSG bins are consistent with the 0.44 ratio, but the highest bin has a $2.33\sigma$ deficit considering statistical error only.

\renewcommand{\arraystretch}{1}

\subsection{\nue survival probability spectrum}\label{sec:spectrum}
\begin{figure}[t]
\centering
\begin{subfigure}{1.0\linewidth}
\includegraphics[width=0.8\linewidth]{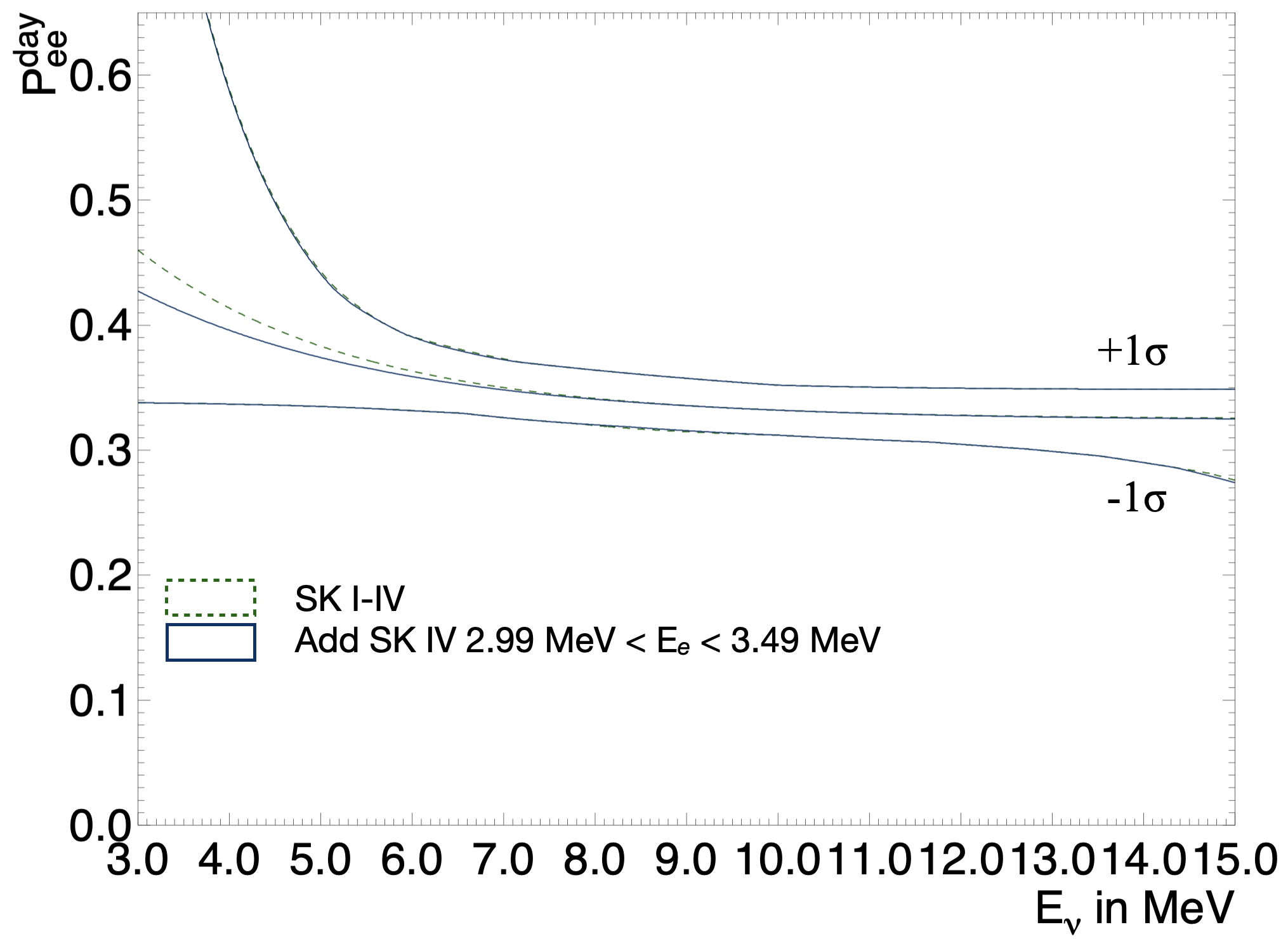}
\caption{Exponential fits.}
\end{subfigure}
\begin{subfigure}{1.0\linewidth}
\includegraphics[width=0.8\linewidth]{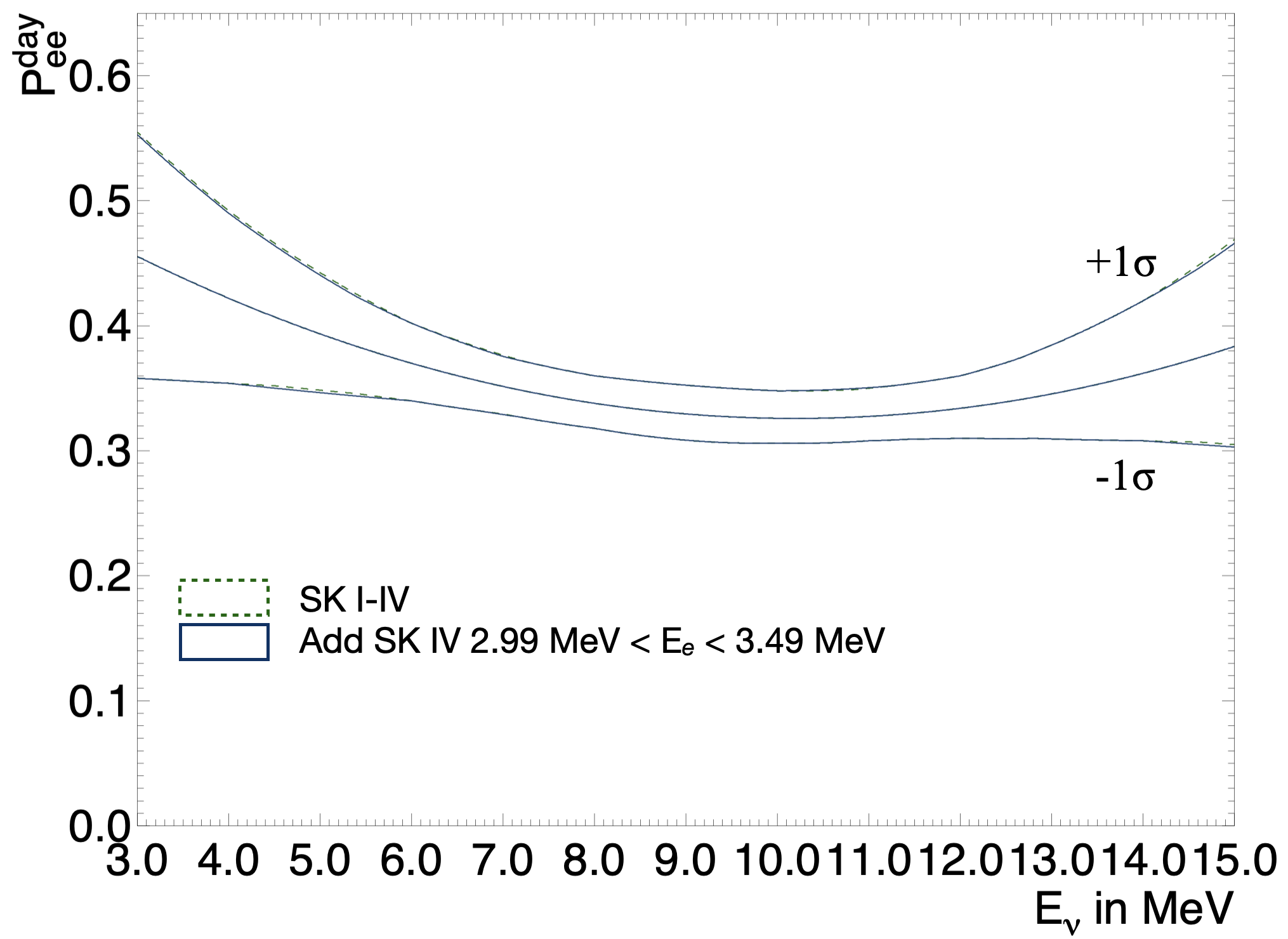}
\caption{Quadratic fits.}
\end{subfigure}
\caption{\label{fig:spec_exp}\nue survival probability spectra as a function of neutrino energy obtained from the exponential (top) and quadratic (bottom) fits to SK recoil electron data. The green dashed lines use all SK-IV and earlier data from \cite{skiv}, and the blue solid lines add an additional \SI{2.99}{MeV}--\SI{3.49}{MeV} bin for SK-IV. The best fits and $1\sigma$ confidence intervals are shown.}
\end{figure}

Following the fits to the solar angle distributions, we repeat the global solar oscillation analysis described in \cite{skiv} after adding the combined \SI{2.99}{MeV}--\SI{3.49}{MeV} measurement in Table \ref{tab:fluxes} to the existing observed-to-expected ratio spectrum of SK-IV. Fits to the \nue survival probability $P_{ee}$ as a function of neutrino energy are then conducted using the quadratic and exponential parametrizations. We then compare these fits along with their $1\sigma$ intervals to the contours of \cite{skiv} in Fig. \ref{fig:spec_exp}.

The additional bin has a low impact on the fits due to the better signal-to-background ratio in the higher-energy bins. The $\chi^2$ of the fit in \cite{skiv} is 65.56 (64.97) for the exponential (quadratic) fit, and these values change to 65.60 (65.03) with the additional bin. There is a slight decrease in the upper confidence interval at lower energies, and there is a noticeable shift in the exponential best fit due to the lower observed to expected measurement compared to higher-energy bins.

\section{Conclusion}
The addition of boosted decision trees for event selection in this analysis of the solar neutrino events during the fourth phase of SK has improved background rejection in the radioactive background-dominated energy region below the previous energy threshold. This analysis utilizes the WIT system implemented during SK-IV, which features improved trigger efficiency at these lower energies and whose data have not yet been used as a source of solar signal events. The results of fits to the solar angle gives an indication for a solar neutrino signal in \SI{2.99}{MeV}--\SI{3.49}{MeV} with a ratio of observed interaction rate to expected without oscillations of $0.307^{+0.091}_{-0.090} \, \rm{(stat.)} \pm 0.066 \, \rm{(syst.)}$. Due to large uncertainties, however, the result is consistent with both the presence and absence of a solar upturn, and the next higher energy bins still provide the bulk of the sensitivity for fits to the \nue survival probability spectrum. The significance and impact of a solar neutrino signal at these energies may improve as selection methods continue to
evolve.

\section*{Acknowledgments}
We gratefully acknowledge the cooperation of the Kamioka Mining and Smelting Company. The Super-Kamiokande experiment has been built and operated from funding by the Japanese Ministry of Education, Culture, Sports, Science, and Technology, the U.S. Department of Energy, and the U.S. National Science Foundation. Some of us have been supported by funds from the National Research Foundation of Korea (NRF-2009-0083526, NRF-2022R1A5A1030700, NRF-2022R1A3B1078756, and RS-2025-00514948) funded by the Ministry of Science, Information, and Communication Technology (ICT), the Institute for Basic Science (IBS-R016-Y2), and the Ministry of Education (2018R1D1A1B07049158, 2021R1I1A1A01042256, and RS-2024-00442775); the Japan Society for the Promotion of Science; the National Natural Science Foundation of China (Grants No. 12375100 and No. 12521007); the Spanish Ministry of Science, Universities, and Innovation (Grant No. PID2021-124050NB-C31); the Natural Sciences and Engineering Research Council (NSERC) of Canada; the Scinet and Digital Research of Alliance Canada; the National Science Centre (UMO-2018/30/E/ST2/00441 and UMO-2022/46/E/ST2/00336) and the Ministry of Science and Higher Education (2023/WK/04), Poland; the Science and Technology Facilities Council (STFC) and Grid for Particle Physics (GridPP), UK; the European Union’s Horizon 2020 Research and Innovation Programme H2020-MSCA-RISE-2018 JENNIFER2 Grant Agreement No. 822070, H2020-MSCA-RISE-2019 SK2HK Grant Agreement No. 872549, and the European Union’s Next Generation EU/PRTR grant CA3/RSUE2021-00559; the National Institute for Nuclear Physics (INFN), Italy.

\section*{Data availability}
The data that support the findings of this article are openly available, including solar angle distributions used to produce Figs. \ref{fig:cossun_wit_msg}--\ref{fig:cossun_sle}, Fig. \ref{fig:cossun_total}, and Table \ref{tab:fluxes}, and the survival probability energy spectra contours for Figure \ref{fig:spec_exp} \cite{data_release}.

\bibliographystyle{apsrev4-2}
\bibliography{bib}



\end{document}

%% file: authors.tex
\newcommand{\AFFicrr}{\affiliation{Kamioka Observatory, Institute for Cosmic Ray Research, University of Tokyo, Kamioka, Gifu 506-1205, Japan}}
\newcommand{\AFFkashiwa}{\affiliation{Research Center for Cosmic Neutrinos, Institute for Cosmic Ray Research, University of Tokyo, Kashiwa, Chiba 277-8582, Japan}}
\newcommand{\AFFipmu}{\affiliation{Kavli Institute for the Physics and
Mathematics of the Universe (WPI), The University of Tokyo Institutes for Advanced Study,
University of Tokyo, Kashiwa, Chiba 277-8583, Japan }}
\newcommand{\AFFmad}{\affiliation{Department of Theoretical Physics, University Autonoma Madrid, 28049 Madrid, Spain}}
\newcommand{\AFFubc}{\affiliation{Department of Physics and Astronomy, University of British Columbia, Vancouver, BC, V6T1Z4, Canada}}
\newcommand{\AFFbu}{\affiliation{Department of Physics, Boston University, Boston, MA 02215, USA}}
\newcommand{\AFFuci}{\affiliation{Department of Physics and Astronomy, University of California, Irvine, Irvine, CA 92697-4575, USA }}
\newcommand{\AFFcsu}{\affiliation{Department of Physics, California State University, Dominguez Hills, Carson, CA 90747, USA}}
\newcommand{\AFFcnm}{\affiliation{Institute for Universe and Elementary Particles, Chonnam National University, Gwangju 61186, Korea}}
\newcommand{\AFFduke}{\affiliation{Department of Physics, Duke University, Durham NC 27708, USA}}
\newcommand{\AFFgifu}{\affiliation{Department of Physics, Gifu University, Gifu, Gifu 501-1193, Japan}}
\newcommand{\AFFgist}{\affiliation{GIST College, Gwangju Institute of Science and Technology, Gwangju 500-712, Korea}}
\newcommand{\AFFuh}{\affiliation{Department of Physics and Astronomy, University of Hawaii, Honolulu, HI 96822, USA}}
\newcommand{\AFFicl}{\affiliation{Department of Physics, Imperial College London , London, SW7 2AZ, United Kingdom }}
\newcommand{\AFFkek}{\affiliation{High Energy Accelerator Research Organization (KEK), Tsukuba, Ibaraki 305-0801, Japan }}
\newcommand{\AFFkobe}{\affiliation{Department of Physics, Kobe University, Kobe, Hyogo 657-8501, Japan}}
\newcommand{\AFFkyoto}{\affiliation{Department of Physics, Kyoto University, Kyoto, Kyoto 606-8502, Japan}}
\newcommand{\AFFliv}{\affiliation{Department of Physics, University of Liverpool, Liverpool, L69 7ZE, United Kingdom}}
\newcommand{\AFFmiyagi}{\affiliation{Department of Physics, Miyagi University of Education, Sendai, Miyagi 980-0845, Japan}}
\newcommand{\AFFnagoya}{\affiliation{Institute for Space-Earth Environmental Research, Nagoya University, Nagoya, Aichi 464-8602, Japan}}
\newcommand{\AFFkmi}{\affiliation{Kobayashi-Maskawa Institute for the Origin of Particles and the Universe, Nagoya University, Nagoya, Aichi 464-8602, Japan}}
\newcommand{\AFFpol}{\affiliation{National Centre For Nuclear Research, 02-093 Warsaw, Poland}}
\newcommand{\AFFsuny}{\affiliation{Department of Physics and Astronomy, State University of New York at Stony Brook, NY 11794-3800, USA}}
\newcommand{\AFFokayama}{\affiliation{Department of Physics, Okayama University, Okayama, Okayama 700-8530, Japan }}
\newcommand{\AFFosaka}{\affiliation{Department of Physics, Osaka University, Toyonaka, Osaka 560-0043, Japan}}
\newcommand{\AFFox}{\affiliation{Department of Physics, Oxford University, Oxford, OX1 3PU, United Kingdom}}
\newcommand{\AFFqmul}{\affiliation{School of Physics and Astronomy, Queen Mary University of London, London, E1 4NS, United Kingdom}}
\newcommand{\AFFregina}{\affiliation{Department of Physics, University of Regina, 3737 Wascana Parkway, Regina, SK, S4SOA2, Canada}}
\newcommand{\AFFseoul}{\affiliation{Department of Physics and Astronomy, Seoul National University, Seoul 151-742, Korea}}
\newcommand{\AFFsheff}{\affiliation{School of Mathematical and Physical Sciences, University of Sheffield, S3 7RH, Sheffield, United Kingdom}}
\newcommand{\AFFshizuokasc}{\affiliation{Department of Informatics in
Social Welfare, Shizuoka University of Welfare, Yaizu, Shizuoka, 425-8611, Japan}}
\newcommand{\AFFstfc}{\affiliation{STFC, Rutherford Appleton Laboratory, Harwell Oxford, and Daresbury Laboratory, Warrington, OX11 0QX, United Kingdom}}
\newcommand{\AFFskk}{\affiliation{Department of Physics, Sungkyunkwan University, Suwon 440-746, Korea}}
\newcommand{\AFFtodai}{\affiliation{Department of Physics, University of Tokyo, Bunkyo, Tokyo 113-0033, Japan }}
\newcommand{\AFFtit}{\affiliation{Department of Physics, Institute of Science Tokyo, Meguro, Tokyo 152-8551, Japan }}
\newcommand{\AFFtus}{\affiliation{Department of Physics and Astronomy, Faculty of Science and Technology, Tokyo University of Science, Noda, Chiba 278-8510, Japan }}
\newcommand{\AFFtriumf}{\affiliation{TRIUMF, 4004 Wesbrook Mall, Vancouver, BC, V6T2A3, Canada }}
\newcommand{\AFFtokai}{\affiliation{Department of Physics, Tokai University, Hiratsuka, Kanagawa 259-1292, Japan}}
\newcommand{\AFFtsinghua}{\affiliation{Department of Engineering Physics, Tsinghua University, Beijing, 100084, China}}
\newcommand{\AFFynu}{\affiliation{Department of Physics, Yokohama National University, Yokohama, Kanagawa, 240-8501, Japan}}
\newcommand{\AFFllr}{\affiliation{Ecole Polytechnique, IN2P3-CNRS, Laboratoire Leprince-Ringuet, F-91120 Palaiseau, France }}
\newcommand{\AFFbari}{\affiliation{ Dipartimento Interuniversitario di Fisica, INFN Sezione di Bari and Universit\`a e Politecnico di Bari, I-70125, Bari, Italy}}
\newcommand{\AFFnapoli}{\affiliation{Dipartimento di Fisica, INFN Sezione di Napoli and Universit\`a di Napoli, I-80126, Napoli, Italy}}
\newcommand{\AFFroma}{\affiliation{INFN Sezione di Roma and Universit\`a di Roma ``La Sapienza'', I-00185, Roma, Italy}}
\newcommand{\AFFpadova}{\affiliation{Dipartimento di Fisica, INFN Sezione di Padova and Universit\`a di Padova, I-35131, Padova, Italy}}
\newcommand{\AFFkeio}{\affiliation{Department of Physics, Keio University, Yokohama, Kanagawa, 223-8522, Japan}}
\newcommand{\AFFwinnipeg}{\affiliation{Department of Physics, University of Winnipeg, MB R3J 3L8, Canada }}
\newcommand{\AFFkcl}{\affiliation{Department of Physics, King's College London, London, WC2R 2LS, UK }}
\newcommand{\AFFwarwick}{\affiliation{Department of Physics, University of Warwick, Coventry, CV4 7AL, UK }}
\newcommand{\AFFral}{\affiliation{Rutherford Appleton Laboratory, Harwell, Oxford, OX11 0QX, UK }}
\newcommand{\AFFwu}{\affiliation{Faculty of Physics, University of Warsaw, Warsaw, 02-093, Poland }}
\newcommand{\AFFbcit}{\affiliation{Department of Physics, British Columbia Institute of Technology, Burnaby, BC, V5G 3H2, Canada }}
\newcommand{\AFFtohoku}{\affiliation{Department of Physics, Faculty of Science, Tohoku University, Sendai, Miyagi, 980-8578, Japan }}
\newcommand{\AFFicise}{\affiliation{Institute For Interdisciplinary Research in Science and Education, ICISE, Quy Nhon, 55121, Vietnam }}
\newcommand{\AFFilance}{\affiliation{ILANCE, CNRS - University of Tokyo International Research Laboratory, Kashiwa, Chiba 277-8582, Japan}}
\newcommand{\AFFibs}{\affiliation{Center for Underground Physics, Institute for Basic Science (IBS), Daejeon, 34126, Korea}}
\newcommand{\AFFglasgow}{\affiliation{School of Physics and Astronomy, University of Glasgow, Glasgow, Scotland, G12 8QQ, United Kingdom}}
\newcommand{\AFFoecu}{\affiliation{Media Communication Center, Osaka Electro-Communication University, Neyagawa, Osaka, 572-8530, Japan}}
\newcommand{\AFFminn}{\affiliation{School of Physics and Astronomy, University of Minnesota, Minneapolis, MN  55455, USA}}
\newcommand{\AFFsilesia}{\affiliation{August Che\l{}kowski Institute of Physics, University of Silesia in Katowice, 75 Pu\l{}ku Piechoty 1, 41-500 Chorz\'{o}w, Poland}}
\newcommand{\AFFtoyama}{\affiliation{Faculty of Science, University of Toyama, Toyama City, Toyama 930-8555, Japan}}
\newcommand{\AFFbmcc}{\affiliation{Science Department, Borough of Manhattan Community College / City University of New York, New York, New York, 1007, USA.}}
\newcommand{\AFFnumazu}{\affiliation{National Institute of Technology, Numazu College, Numazu, Shizuoka 410-8501, Japan}}

\AFFicrr
\AFFkashiwa
\AFFmad
\AFFbmcc
\AFFbu
\AFFbcit
\AFFuci
\AFFcsu
\AFFcnm
\AFFduke
\AFFllr
\AFFgifu
\AFFgist
\AFFglasgow
\AFFuh
\AFFibs
\AFFicise
\AFFicl
\AFFbari
\AFFnapoli
\AFFpadova
\AFFroma
\AFFilance
\AFFkeio
\AFFkek
\AFFkcl
\AFFkobe
\AFFkyoto
\AFFliv
\AFFminn
\AFFmiyagi
\AFFnagoya
\AFFkmi
\AFFpol
\AFFnumazu
\AFFsuny
\AFFokayama
\AFFoecu
\AFFox
\AFFral
\AFFseoul
\AFFsheff
\AFFshizuokasc
\AFFsilesia
\AFFstfc
\AFFskk
\AFFtohoku
\AFFtodai
\AFFipmu
\AFFtit
\AFFtus
\AFFtoyama
\AFFtriumf
\AFFtsinghua
\AFFwu
\AFFwarwick
\AFFwinnipeg
\AFFynu

\author{A.~Yankelevich}
\AFFuci

\author{K.~Abe}
\AFFicrr
\AFFipmu
\author{Y.~Asaoka}
\AFFicrr
\AFFipmu
\author{M.~Harada}
\AFFicrr
\author{Y.~Hayato}
\AFFicrr
\AFFipmu
\author{K.~Hiraide}
\AFFicrr
\AFFipmu
\author{T.~H.~Hung}
\AFFicrr
\author{K.~Hosokawa}
\AFFicrr
\author{K.~Ieki}
\author{M.~Ikeda}
\AFFicrr
\AFFipmu
% ...


\collaboration{The Super-Kamiokande Collaboration}
\noaffiliation

%% file: commands.tex
\newcommand{\nue}{\ensuremath{\nu_{e}}\xspace}
\newcommand{\nuecc}{\ensuremath{\nu_{e}}~CC\xspace}
\newcommand{\nuebar}{\ensuremath{\overbar{\nu}_e}\xspace}
\newcommand{\nuebarcc}{\ensuremath{\overbar{\nu}_e}~\text{CC}\xspace}
\newcommand{\nueParen}{\ensuremath{{\nu}_{e}~(\bar{\nu}_{e})}\xspace}
\newcommand{\nueParenCC}{\ensuremath{{\nu_{e}~(\bar{\nu}_{e})}}~\text{CC}\xspace}
\newcommand{\antinue}{\nuebar}
\newcommand{\numu}{\ensuremath{\nu_{\mu}}\xspace}
\newcommand{\numucc}{\ensuremath{\nu_{\mu}}~\text{CC}\xspace}
\newcommand{\numubar}{\ensuremath{\overbar{\nu}_\mu}\xspace}
\newcommand{\numubarcc}{\ensuremath{\overbar{\nu}_\mu}~\text{CC}\xspace}
\newcommand{\numuParen}{\ensuremath{{\nu}_{\mu}~(\bar{\nu}_{\mu})}\xspace}
\newcommand{\numuParenCC}{\ensuremath{{\nu_{\mu}~(\bar{\nu}_{\mu})}}~\text{CC}\xspace}
\newcommand{\antinumu}{\numubar}
\newcommand{\nutau}{\ensuremath{\nu_{\tau}}\xspace}
\newcommand{\nutaubar}{\ensuremath{\overbar{\nu_\tau}}\xspace}

\newcommand{\nuone}{\ensuremath{\nu_{1}}\xspace}
\newcommand{\nuonebar}{\ensuremath{\bar{\nu}_{1}}\xspace}
\newcommand{\nutwo}{\ensuremath{\nu_{2}}\xspace}
\newcommand{\nutwobar}{\ensuremath{\bar{\nu}_{2}}\xspace}
\newcommand{\nuthree}{\ensuremath{\nu_{3}}\xspace}
\newcommand{\nuthreebar}{\ensuremath{\bar{\nu}_{3}}\xspace}